\documentclass[a4paper,11pt]{article} 
\pdfoutput=1 
\usepackage{jheppub} 

\usepackage{amsmath,graphicx,hyperref}
\usepackage{xcolor}

\newcommand{\fms}[1]{{#1}\!\!\!/}
\newcommand{\fmsl}[1]{{#1}\!\!\!\!/}

\newcommand{\mc}{\mathcal}
\newcommand{\mr}{\mathrm}

\newcommand{\be}{\begin{equation}} 
\newcommand{\ee}{\end{equation}} 
\newcommand{\bea}{\begin{eqnarray}} 
\newcommand{\eea}{\end{eqnarray}}

\newcommand{\dg}{\dagger}
\newcommand{\n}{\overline{n}}
\newcommand{\nn}{\frac{\fms{\overline{n}}}{2}} 
\newcommand{\nnn}{\frac{\fms{n}}{2}} 

\newcommand{\bl}[1]{{\bf{#1}}}

\newcommand{\blp}[1]{{\bf{#1}}_{\perp}}
\newcommand{\blpu}[1]{{\bf{#1}}^{\perp}}

\newcommand{\bsp}[1]{{\boldsymbol{#1}}_{\perp}}

\newcommand{\nnb}{\nonumber} 

\newcommand{\as}{\alpha_s} 

\newcommand{\eps}{\epsilon} 
\newcommand{\veps}{\varepsilon}

\newcommand{\zc}{z_{\mr{cut}}}

\begin{document}



\title{Heavy quark jet production near threshold}

\def\Seoultech{Institute of Convergence Fundamental Studies and School of Liberal Arts, Seoul National University of Science and Technology, Seoul 01811, Korea}
\def\Pitt{Pittsburgh Particle Physics Astrophysics and Cosmology Center (PITT PACC) \\ Department of Physics and Astronomy, University of Pittsburgh, Pittsburgh, Pennsylvania 15260, USA}
\def\BenGurion{Department of Physics, Ben-Gurion University of the Negev, Beer-Sheva 84105, Israel}

\author[a]{Lin Dai}
\emailAdd{dail@post.bgu.ac.il}
\affiliation[a]{\BenGurion}
\author[b]{Chul Kim}
\emailAdd{chul@seoultech.ac.kr}
\affiliation[b]{\Seoultech} 
\author[c]{Adam K. Leibovich}
\emailAdd{akl2@pitt.edu}
\affiliation[c]{\Pitt}

\abstract{ 
In this paper, we study the fragmentation of a heavy quark into a jet near threshold, meaning that final state jet carries most of the energy of the fragmenting heavy quark. Using the heavy quark fragmentation function,  we simultaneously resum large logarithms of the jet radius $R$ and $1-z$, where $z$ is the ratio of the jet energy to the initiating heavy quark energy. There are numerically significant corrections to the leading order rate due to this resummation. We also investigate the heavy quark fragmentation to a groomed jet, using the soft drop grooming algorithm as an example. In order to do so, we introduce a collinear-ultrasoft mode sensitive to the grooming region determined by the algorithm's $z_{\mathrm{cut}}$ parameter. This allows us to resum large logarithms of $z_{\mathrm{cut}}/(1-z)$, again leading to large numerical corrections near the endpoint. A nice feature of the analysis of the heavy quark fragmenting to a groomed jet is the heavy quark mass $m$ renders the algorithm infrared finite, allowing a perturbative calculation.  We analyze this for $E_JR \sim m$ and $E_JR\gg m$, where $E_J$ is the jet energy. To do the latter case, we introduce an ultracollinear-soft mode, allowing us to resum large logarithms of $E_JR/m$. Finally, as an application we calculate the rate for $e^+e^-$ collisions to produce a heavy quark jet in the endpoint region, where we show that grooming effects have a sizable contribution near the endpoint.

}

\maketitle 


\section{Introduction} 

Jets containing a heavy quark are an important experimental tool for probing many interesting areas of particle physics, including Higgs physics~\cite{Butterworth:2008iy}, top physics~\cite{Fleming:2007qr,Fleming:2007xt,Almeida:2008tp}, and many proposed beyond the Standard Model scenarios. As such, it is important to have a good theoretical understanding of the production and properties of heavy quark jets, 
which have been lately actively studied in various high energy processes~\cite{Bauer:2013bza,Dai:2018ywt,Makris:2018npl,Li:2018xuv,Gauld:2019doc,Lee:2019lge,Lepenik:2019jjk,Kim:2020dgu,Kang:2020xgk,Czakon:2021ohs}. 

Jets are formed by initially producing a very energetic parton, which radiates many collinear and soft particles into a highly collimated beam. By using the concept of a fragmentation function, we can accurately describe the jet formation process~\cite{Dasgupta:2014yra,Kaufmann:2015hma,Kang:2016mcy,Dai:2016hzf,Qiu:2019sfj}, including jets that contain a heavy quark~\cite{Dai:2018ywt}.  Of particular importance both theoretically and experimentally are energetic jets with the heavy quark highly boosted, meaning that the heavy quark carries a large fraction of the total jet energy. In this region of phase space, the amount of radiated energy is restricted, which can lead to large numerical corrections to the calculated rate. 

A jet is defined using a specific jet algorithm in order to define how particles are sorted to be within or outside of the jet. Most jet algorithms use a parameter to differentiate the two sets of particles, usually denoted as the jet radius $R$. We can also define the fraction of the jet energy that is initiated by the heavy quark, $z=E_J/E_Q$. Focusing now on the radius, when $R$ is small, terms including $\ln R$ appear in the production rate calculation that can spoil the QCD perturbative expansion. In order to obtain reliable theoretical predictions, these logarithms must be summed. As shown in Refs.~\cite{Dasgupta:2014yra,Kaufmann:2015hma,Kang:2016mcy,Dai:2016hzf,Qiu:2019sfj}, these logarithms can be resummed by running the scale down to $\mu\sim QR$ using the Dokshitzer-Gribov-Lipatov-Altarelli-Parisi (DGLAP) evolution equations, where $Q$ is the hard scale of order of the jet energy $E_J$. In Ref.~\cite{Dai:2016hzf} we studied the fragmentation of light partons to a jet with a small radius, resumming the large logarithms of $R$, while in Ref.~\cite{Dai:2018ywt}, we investigated the fragmentation of a jet containing a heavy quark, again resumming these large logarithms.

We also have the possible situation where the jet carries a large fraction of the energy that was carried by the initiating parton, i.e., $z\sim 1$. In this region of phase space, we have large logarithms of $\ln(1-z)$, again potentially spoiling the perturbative expansion. Therefore, again for a proper description we need to systematically resum the large logarithms of $1-z$~\cite{Dai:2017dpc,{Liu:2017pbb,Nagy:2017dxh,Liu:2018ktv,Kaufmann:2019ksh,Neill:2021std}}.  Since jets are made up of collinear particles, and in the large $z$ limit only soft particles can be emitted out of the jet, soft-collinear effective theory (SCET) \cite{Bauer:2000ew,Bauer:2000yr,Bauer:2001yt,Bauer:2002nz} is a natural tool to use to facilitate the resummation of these logarithms. 

In order to resum the logarithms of $1-z$, we need to look at the dynamics in the endpoint region. Doing so, we can see that there are a number of distinct, well-separated energy scales. Since the observed jet carries most of the energy of the initiating parton, all radiation that falls outside of the jet must be soft, $\sim E_J(1-z)$. However,  the standard modes in SCET are not adequate for separating the radiation inside and outside of the jet, since they are not sensitive to the jet boundary  characterized by the small radius $R$. We thus must introduce a new soft mode, called the collinear-soft mode \cite{Bauer:2011uc,Procura:2014cba}, which has the ability to resolve the jet boundary and can consistently discriminate between soft radiation that falls inside versus outside of the jet. Using this extra mode, we were able to resum both the large logarithms of $R$ and $1-z$ for a massless parton initiated jet \cite{Dai:2017dpc}. One of the goals of this paper is to extend this calculation to jets initiated by heavy quarks, which can potentially give large numerical corrections.  

The study of jet substructure has developed rapidly in recent years (see Ref.~\cite{Larkoski:2017jix} for a recent review), due to advancement in jet grooming algorithms~\cite{Butterworth:2008iy,Ellis:2009su,Ellis:2009me,Krohn:2009th,Larkoski:2014wba}. These jet grooming algorithms remove soft radiation contamination in the jet by  first ``declustering" a jet and then removing soft radiation to form the new groomed jet, following a precise recipe that depends on some input constants. For instance, the soft drop algorithm~\cite{Larkoski:2014wba} depends on two variables, $\zc$ and $\beta$, which determine if two subjets are to be treated as distinct or one combined jet. Since these jet grooming algorithms are used to probe the substructure of jets, it is important to understand the soft gluon radiations that may or may not fall within the jet. Therefore, we apply our above analysis for when we have a heavy quark fragmenting to groomed jet. 
Compared with similar analyses of jet grooming with a light quark~\cite{Larkoski:2014wba,Larkoski:2015lea,Marzani:2017kqd,Cal:2020flh}, the heavy quark fragmentation is perfectly free from infrared (IR) divergence due to the nonzero quark mass, but the factorization and its calculation becomes more intricate and interesting.
However, as it does for the light quark jet, the grooming can cause large numerical corrections in the endpoint region.

The organization of this paper is as follows. In Sec.~\ref{factorization} we investigate the factorization of the heavy quark fragmentation function to a jet (FFJ) in the large $z$ limit.  To do so, we will be using an inclusive $\mr{k_T}$-type jet algorithm~\cite{Catani:1993hr,Ellis:1993tq,Dokshitzer:1997in,Cacciari:2008gp}. We need to introduce the collinear-soft mode, as described above, and we implement boosted heavy quark effective theory to describe the interactions of the heavy quark with these collinear-soft interactions. (Details about the construction of boosted heavy quark effective theory can be found in Appendix~\ref{bHQET}.)  In Sec.~\ref{NLOFFJ} we calculate the next-to-leading order heavy quark FFJ in the large $z$ limit, resumming the large logarithms of $R$ and $1-z$.  In Sec.~\ref{sec4} we investigate the fragmentation of a  heavy quark fragmenting to a groomed jet. This allows us, in Sec.~\ref{sec5}, to combine our results above to calculate the heavy quark fragmentation function to a groomed jet, again resumming the large logarithms involved. As an example, we calculate the $e^+e^-$ cross section to a heavy quark  jet near threshold, comparing the groomed and ungroomed results. Finally, we conclude in Sec.~\ref{conclusion}. There is also a brief discussion of profile functions in Appendix~\ref{sec:profile}.

\section{Factorization of the Heavy Quark FFJ in the large $z$ limit}
\label{factorization}

If we consider a jet with a small radius~($R$), the relevant dynamics are in general described by collinear interactions. Thus, the fragmenting process to a jet from a heavy quark can be properly studied using the framework of SCET, more specifically $\mr{SCET_M}$~\cite{Leibovich:2003jd,Rothstein:2003wh,Chay:2005ck}. For this paper, we will consider the case $E_J \gg m$, but for now to be as generic as possible we will not set a hierarchy between $R$ and $m/E_J$, where $m$ is the heavy quark mass. 
Using SCET, we can express the fragmentation function to a jet (FFJ) initiated by a heavy quark in $D~(=4-2\eps)$~dimension as \cite{Dai:2018ywt}
\begin{align}
D_{J/Q}(z;E_JR',m,\mu) &= \sum_{X}\frac{1}{2N_cz}  \int d^{D-2}\blpu{p}_J  \mr{Tr} \langle 0 | \delta \Bigl(\frac{p_J^+}{z}-\mc{P}_+\Bigr)\delta^{(D-2)}(\bsp{\mc{P}})\nn \Psi_n^Q  \nnb\\ 
\label{defHQFFJ} 
&~~~~~~~~~~~\times| J(p_J^+, \blpu{p}_J) X_{\notin J}\rangle
\langle J(p_J^+, \blpu{p}_J) X_{\notin J} | \bar{\Psi}_n^Q |0\rangle,
\end{align} 
where $N_c$ is the number of colors and $Q$ denotes the heavy quark. $X_{\notin J}$ are the final states that are not contained within the observed jet $J$. 
$p_J$ is the momentum of the jet $J$, which can be depicted as an $n$-collinear object. This means the jet momentum can be decomposed and power-counted as 
\be
p_J^{\mu} = (\n\cdot p_J, \blpu{p}_J, n\cdot p) = (p_J^+, \blpu{p}_J, p_J^-) \sim E_J (1, R, R^2), 
\ee
where $n$ and $\n$ are the lightcone vectors normalized to $n\cdot \n =2$, and $p_J^+$ is approximately $p_J^+\approx2E_J$. 
$\Psi_n^Q = W_n^{\dagger} \xi_n^Q$ is the gauge invariant massive collinear field, where $W_n$ is the collinear Wilson line. 

When we calculate the heavy quark FFJ at one loop, we will employ an inclusive $\mr{k_T}$-type jet algorithm, which at this order covers  
$\mr{k_T}$~\cite{Catani:1993hr,Ellis:1993tq}, C/A~\cite{Dokshitzer:1997in}, and anti-$\mr{k_T}$~\cite{Cacciari:2008gp} algorithms. Under the algorithm, when two particles are merged into a jet at one loop, their opening angle $\theta$ should satisfy 
\be
\label{jetmerg}
\theta < R'.
\ee 
Here $R'=R$ for $e^+e^-$ annihilation, and $R'=R/\cosh{y}$ for hadron collision, where $y$ is the rapidity of the jet relative to the hadron beam direction. For a parton splitting with momenta $q \to p +k$, the phase space constraint from Eq.~\eqref{jetmerg} is  
\bea
\label{pconq}
\tan^2 \frac{R'}{2} &>& \frac{q_+^2\blp{k}^2}{p_+^2k_+^2},~~~~~~(\blp{q} = 0),\\
\label{pconp}
\tan ^2\frac{R'}{2} &>& \frac{\blp{k}^2}{k_+^2},~~~~~~~~~(\blp{p} = 0),
\eea
depending on which frame we choose. These constraints apply not only to massless partons, but also to massive partons as long as the parton masses are much smaller than their energies. 

The splitting and fragmenting processes in Eq.~\eqref{defHQFFJ}, have been studied in Ref.~\cite{Dai:2018ywt} for the full range of $z$. At one loop the splitting of a heavy quark to a jet is divided into two processes, $Q \to J_Q + g$ and $Q \to J_g+Q$, where $J_i$ denotes the jet containing the parton $i$. The renormalization behavior for each process satisfies the well known DGLAP evolution. 

If we focus on the region where $z$ is close to 1, the process $Q \to J_g +Q$ is suppressed by $1-z$. Therefore the fragmenting function can be described  by the dominant process $Q \to J_Q +g$. In this case the jet $J_Q$ takes most of the energy of the mother parton $Q$, while radiation out of the jet will be carried by gluons with energy $\sim E_J(1-z)$. 
Furthermore, these gluons need to be sensitive to the jet boundary restricted by small jet radius $R$. 
Thus in SCET these gluons can be considered to be quanta of a collinear-soft (csoft) field~\cite{Bauer:2011uc,Procura:2014cba}, with momenta schematically given by
\be
p_{cs} = (p_{cs}^+,p_{cs}^{\perp}, p_{cs}^-) \sim (1-z) E_J (1,R,R^2). 
\ee

Therefore, in order to properly describe the heavy quark FFJ in the large $z$ limit, we need to separate csoft interactions from collinear interactions. Note that collinear parton radiations are still allowed inside the jet, but radiations outside the jet are carried out solely by csoft gluons.  
As a result, similar to with the massless parton case~\cite{Dai:2017dpc}, the heavy quark FFJ can be factorized into the collinear and the csoft parts as $z$ goes to 1. 

\subsection{Integrated heavy quark jet function}
\label{secIHQJF}

Since the collinear radiations reside only inside the jet, their contribution to the heavy quark FFJ can be obtained from integrating the collinear interactions out to the jet boundary in the phase space. 
We will call it the integrated heavy quark jet function (iHQJF), for which the detailed one-loop calculation can be found in Ref.~\cite{Dai:2018ywt}. The bare one-loop result is 
\bea
\label{IHQJF} 
\mc{J}_Q (E_JR',m,\mu) &=& 1+ \frac{\as C_F}{2\pi} \Biggl[\frac{1}{\eps^2}+\frac{1}{\eps} \Bigl(\ln \frac{\mu^2}{E_J^2 R'^2+m^2}+\frac{3+b}{2(1+b)}\Bigr) +\frac{1}{2} \ln\frac{\mu^2}{E_J^2 R'^2+m^2} \nnb \\
&&~~+ \frac{1}{1+b}\Bigl(\ln \frac{\mu^2}{E_J^2 R'^2} +2\Bigr) +\frac{1}{2} \ln^2 \frac{\mu^2}{E_J^2 R'^2+m^2} -\frac{1}{2} \ln^2 (1+b) \nnb \\
&&~~+ f(b)+g(b)-\mr{Li}_2 (-b) + 2 -\frac{\pi^2}{12} \Biggr].
\eea
Here the $1/\eps$ poles are all ultraviolet (UV) divergences, $b \equiv m^2/(E_J^2R'^2)$, and the functions $f(b)$ and $g(b)$ are defined by the following integrals: 
\bea
\label{fb}
f(b) &=& \int^1_0 dz \frac{1+z^2}{1-z} \ln\frac{z^2+b}{1+b}, \\
\label{gb}
g(b) &=& \int^1_0 dz \frac{2z}{1-z}\Bigl(\frac{1}{1+b}-\frac{z^2}{z^2+b}\Bigr). 
\eea
If we take the limit $m\to 0$, Eq.~\eqref{IHQJF} reduces to the result for a massless quark \cite{Cheung:2009sg,Ellis:2010rwa,Liu:2012sz,Chay:2015ila}  
\bea
\label{IJF} 
\mc{J}_q (E_JR',\mu) &=& 1+ \frac{\as C_F}{2\pi} \Biggl[\frac{1}{\eps^2}+\frac{1}{\eps} \Bigl(\ln \frac{\mu^2}{E_J^2 R'^2}+\frac{3}{2}\Bigr) +\frac{3}{2} \ln\frac{\mu^2}{E_J^2 R'^2} \nnb \\
&&\phantom{1+ \frac{\as C_F}{2\pi} \Biggl[}+\frac{1}{2} \ln^2 \frac{\mu^2}{E_J^2 R'^2} +\frac{13}{2} -\frac{3\pi^2}{4}\Biggr].  
\eea

Comparing $\mc{J}_Q$ with the massless case $\mc{J}_q$, one may wonder how the UV poles of $\mc{J}_Q$ in Eq.~\eqref{IHQJF} could depend on the heavy quark mass. Usually a finite quark mass in the amplitude would be irrelevant as the energy or momentum transfer becomes infinite. In order to understand this, we need to keep in mind that the integrated jet functions can only be properly obtained after the zero-bin subtraction~\cite{Manohar:2006nz}.  In the naive collinear calculation for $\mc{J}_Q$, the poles that depend on the heavy quark mass are present as infrared (IR) divergences. However, when contributions from the zero-bin mode are subtracted to avoid double counting, these IR poles are cancelled and  are all converted to UV poles due to the so-called pull-up mechanism. See Ref.~\cite{Manohar:2006nz} for complete details. 

\begin{figure}[t]
\begin{center}
\includegraphics[height=4cm]{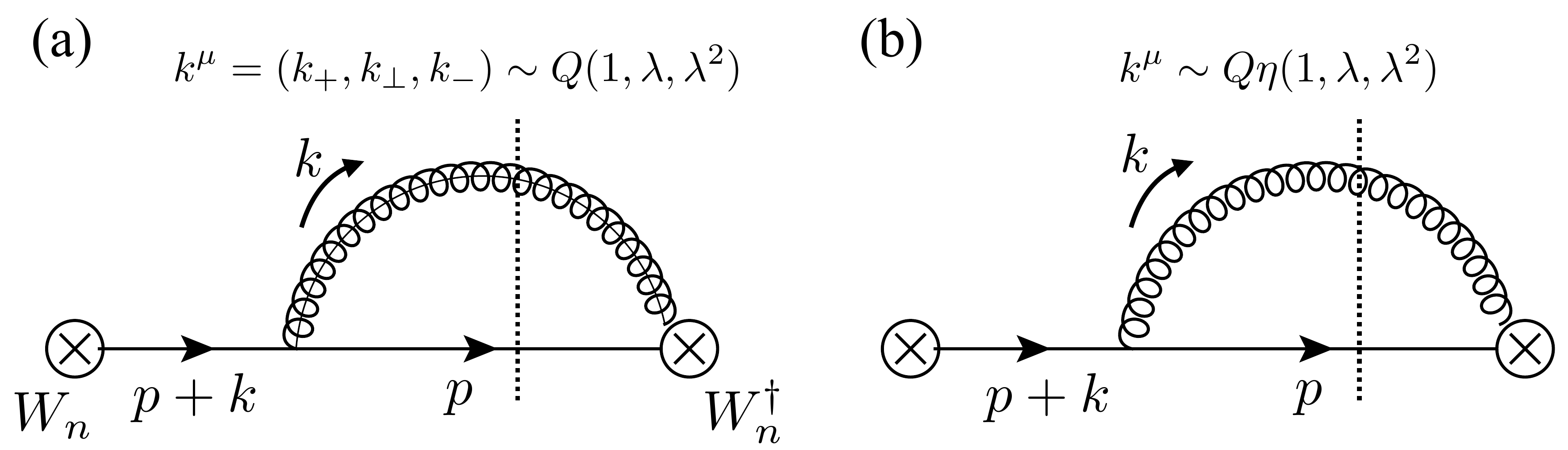}
\end{center}
\vspace{-0.8cm}
\caption{\label{fig1}\baselineskip 3.5ex
(a) One sample Feynman diagram for real gluon radiation contributing to the integrated heavy quark jet function. Diagram (b) corresponds to the zero-bin contribution to the diagram (a). Dotted lines indicate unitary cuts where particles are on-shell.   }
\end{figure}

In Fig.~\ref{fig1}~(a), we show an example Feynman diagram that contributes to the real radiation in calculating iHQJF.  
The gluon coming from the heavy quark radiates inside the jet and has collinear scaling:  $k^{\mu} \sim Q(1,\lambda,\lambda^2)$, where $Q \sim E_J$ and $\lambda \sim R \sim m/Q$. We see that iHQJF is obtained from integrating out the collinear radiation and it can be regarded as a Wilson coefficient when matched onto the lower energy effective theory with csoft interactions. Therefore, in order to obtain the correct contribution to iHQJF from Fig.~\ref{fig1}~(a), we need to subtract the zero-bin contribution of Fig.~\ref{fig1}~(b), where the gluon carries a csoft momentum scaling $k^{\mu} \sim Q\eta(1,\lambda,\lambda^2)$. Here $\eta$ is a new small parameter comparable to $1-z$ when we investigate the threshold region in the large $z$ limit. 

To see the divergence structure of Fig.~\ref{fig1}~(a), we compute 
\bea
\mathcal{M}_a &=& 2\pi g_s^2 C_F \mu_{\overline{MS}}^{2\eps} \int^1_0 \frac{dx}{1-x} \int_{\Omega_\mr{In}} \frac{d^D k}{(2\pi)^D} \delta\Bigl(\frac{x}{1-x} p_+ - k_+\Bigr) \frac{(p+k)_+}{(p+k)^2-m^2} \frac{2p_+}{k_+} \delta(k^2) \nnb \\
\label{Rncol}
&=& \frac{\as C_F}{2\pi} \frac{(\mu^2 e^{\gamma_E})^{\eps}}{\Gamma(1-\eps)} \int^1_0 dx\frac{1-x}{x} 
\int^{x^2(1-x)^2 E_J^2 R'^2}_0 \frac{d\blp{k}^2(\blp{k}^2)^{-\eps}}{\blp{k}^2+x^2m^2},
\eea
where $4\pi\mu_{\overline{MS}}^2 = \mu^2 e^{\gamma_E}$, and $p$ and $k$ are on-shell momenta of the heavy quark and the gluon, respectively. $x$ is the large momentum fraction of the gluon over the jet momentum, $x = k_+/(p+k)_+ = k_+/(2E_J)$. In the first equality of Eq.~\eqref{Rncol}, 
$\Omega_\mr{In}$ represents the phase space inside a jet. We have chosen the frame $\bl{p}_{J\perp} = \blp{p}+\blp{k}=0$, so we employed the phase space constraint shown in Eq.~\eqref{pconq}. As can be seen from the second equality of Eq.~\eqref{Rncol}, the collinear IR divergence from $\blp{k}^2 \to 0$ does not appear due to the presence of the heavy quark mass. However, the soft IR divergence that happens as $x\to 0$ is still presents and is regularized in dimensional regularization as a pole in $\eps$,  
\be
\label{Rncolr}
\mathcal{M}_a = \frac{\as C_F}{2\pi} \Biggl[-\frac{1}{2} \ln \frac{1+b}{b}\Bigl(\frac{1}{\eps_{\mr{IR}}} - \ln \frac{\mu^2}{E_J^2 R'^2}\Bigr) + \cdots\Biggr],
\ee
where we suppressed other finite terms. It makes sense that the pole is IR since the energy of the collinear gluon is limited by the jet energy and cannot grow infinitely large. 

When considering the zero-bin contribution, notice that the radiated csoft gluon does not depend on the jet energy. Hence, 
within the zero-bin, its energy is unlimited and can freely grow to infinity, although it can be regarded to be much smaller than the heavy quark energy. Taking these into account, we compute the diagram of Fig.~\ref{fig1}~(b),  
\bea
\mc{M}_b &=& 2\pi g_s^2 C_F \mu_{\overline{MS}}^{2\eps} \int^{\infty}_0 dx 
\int_{\Omega_\mr{In}} \frac{d^D k}{(2\pi)^D} \delta(x p_+ - k_+) \frac{p_+}{(p+k)^2-m^2}\frac{2p_+}{k_+} \delta(k^2) \nnb \\
\label{Rzcsoft}
&=& \frac{\as C_F}{2\pi} \frac{(\mu^2 e^{\gamma_E})^{\eps}}{\Gamma(1-\eps)} \int^{\infty}_0 \frac{dx}{x} 
\int^{x^2 E_J^2 R'^2}_0 \frac{d\blp{k}^2(\blp{k}^2)^{-\eps}}{\blp{k}^2+x^2m^2}.
\eea
Here the denominator of the heavy quark propagator with off-shell momentum $p+k$ has been approximated as\footnote{\baselineskip 3.0ex
For real gluon radiation, the relation, $(p+k)^2 -m^2 = 2p\cdot k$, is satisfied even if $k$ is collinear. However, for the virtual radiation, the relation (approximately) holds only when $k$ is a csoft momentum. So, as long as $k$ is csoft, this relation always holds, whether $k$ is on-shell or not. 
}   
\be
\label{mappr}
(p+k)^2-m^2 = k^2 + 2p\cdot k \approx 2p\cdot k,
\ee
since $k^2 \sim Q^2\eta^2 \lambda^2$ is power-suppressed by $\eta$ when compared with $2p\cdot k$. 
For the real radiation shown in Fig.~\ref{fig1}~(b), Eq.~\eqref{mappr} is expressed as 
\be
\label{mappr2}
2p\cdot k \approx p_+ k_- + p_-k_+ = \frac{1}{x} (\blp{k}^2 + x^2 m^2),
\ee
where we have used $\blp{p} = 0$ due to $\bl{p}_{J\perp} = 0 =  \blp{p}+\blp{k} \sim \blp{p}$ in our chosen frame. Then we use the on-shell conditions: $p_- = m^2/p_+$ and $k_- = \blp{k}^2/k_+$. In the power counting of Eq.~\eqref{mappr2}, $\blp{k}^2$ and $x^2 m^2$ are  the same order, $\mc{O}(\eta^2 \lambda^2)$. 

Finally, Eq.~\eqref{Rzcsoft} has the result 
\be
\label{rMb}
\mc{M}_b  = \frac{\as C_F}{4\pi} \ln \frac{1+b}{b} \Bigl(\frac{1}{\eps_{\mr{UV}}} - \frac{1}{\eps_{\mr{IR}}}\Bigr). 
\ee
In obtaining Eq.~\eqref{rMb}, we separated UV and IR poles in the integration of $x$ as
\be
\int^{\infty}_0 dx~x^{-1-\eps} = \frac{1}{\eps_{\mr{UV}}} - \frac{1}{\eps_{\mr{IR}}}. 
\ee
Therefore, while the real contribution of Fig.~\ref{fig1} was originally IR-divergent, the pole can be converted to be UV after the zero-bin subtraction:   
\be
\label{Rcolla}
\mathcal{M}_a -\mathcal{M}_b = \frac{\as C_F}{2\pi} \Biggl[-\frac{1}{2} \ln \frac{1+b}{b}\Bigl(\frac{1}{\eps_{\mr{UV}}} - \ln \frac{\mu^2}{E_J^2 R'^2}\Bigr) + \cdots\Biggr]. 
\ee
Similar conversions by the zero-bin subtraction works for the other real radiation as well as the virtual contributions.  
As a result we obtain Eq.~\eqref{IHQJF}, which only has genuinely UV divergences. 

\subsection{Description of the heavy quark csoft interactions using  boosted heavy quark effective theory} 

In Sec.~\ref{secIHQJF} we saw that the iHQJF is the result of integrating out collinear interactions of the heavy quark and the matching coefficient onto the lower energy theory with csoft interactions. The zero-bin contribution to iHQJF that needs to be subtracted can be interpreted as a contribution contained in the lower energy effective theory. As seen from Eq.~\eqref{mappr}, when the (boosted) heavy quark interacts with a csoft gluon, the denominator of the heavy quark propagator can be approximated as $2p\cdot k$. In this case the on-shell heavy quark momentum $p$ can be written as $mv$ since the heavy quark velocity $v$ does not change under the csoft interaction. Therefore, the denominator is proportional to $v\cdot k$, and it indicates that the lower energy effective theory for the boosted heavy quark with csoft interactions should be the boosted heavy quark effective theory (bHQET).   

Since we describe the boosted heavy quark as an $n$-collinear particle, the velocity $v$ is also $n$-collinear. In general, whether the heavy quark is on-shell or not, the momentum under csoft interactions can be written as 
\be
p^{\mu} = mv^{\mu} + k^{\mu}, 
\ee
where $k^{\mu}$ is a residual csoft momentum. The velocity $v$ has scaling $v^{\mu} = (v_+,v_{\perp},v_-) \sim (1/\lambda,1,\lambda)$, where $\lambda \sim m/p_+$. If we choose the frame with $v_{\perp}=0$, the velocity $v$ can be simply written as 
\be
\label{velo} 
v^{\mu} = v_+ \frac{n^{\mu}}{2} + v_- \frac{\n^{\mu}}{2} = v_+ \frac{n^{\mu}}{2} + \frac{1}{v_+} \frac{\n^{\mu}}{2}. 
\ee 
In the second equality, $v_-$ has been rewritten as $1/v_+$ using $v^2 = v_+v_-=1$. 

To describe collinear interactions of the heavy quark, we already employed the effective theory $\mr{SCET_M}$. 
We thus want to directly match $\mr{SCET_M}$ onto bHEQT, integrating out collinear interactions. For this, we match the collinear quark field in $\mr{SCET_M}$ onto the bHQET field,
\be
\label{Qmatch}
\xi_n (x) = \sqrt{\frac{v_+}{2}} e^{-imv\cdot x} h_n (x). 
\ee
Note that the bHQET field $h_n$ shares the same spinor property as $\xi_n$, satisfying 
\be 
\fms{n} h_n =0,~~\frac{\fms{n}\fms{\n}}{4} h_n = h_n.
\ee
This has the advantage that the power counting on the large energy in $\mr{SCET_M}$ remains true at the lower scale.    

In $\mr{SCET_M}$, if we separate csoft interactions from collinear interactions, the covariant derivative needs to be written as 
\be 
\label{deriv}
i\mc{D}^{\mu} = iD_c^{\mu} + iD_{cs}^{\mu}, 
\ee
where the csoft derivative $iD_{cs}^{\mu}$ is suppressed by $\eta$, and thus does not give a leading contribution in $\mr{SCET_M}$. At an  energy scale lower than a typical scale for $\mr{SCET_M}$, collinear gluons are not present, and  $iD_c^{\mu}$ in Eq.~\eqref{deriv} becomes the derivative operator $\mc{P}^{\mu}$ that picks up the frozen collinear heavy quark momentum,
\be 
\label{deriv2}
i\mc{D}^{\mu} \to \mc{P}^{\mu} + iD_{cs}^{\mu}.
\ee  
In constructing bHQET from the $\mr{SCET_M}$ Lagrangian using Eq.~\eqref{Qmatch}, $\mc{P}^{\mu}$ returns $mv^{\mu}$, and hence $i\mc{D}^{\mu}$ in Eq.~\eqref{deriv} can be expressed as $mv^{\mu} + iD_{cs}^{\mu}$. The details of constructing the bHQET Lagrangian is shown in  Appendix~\ref{bHQET}. The leading bHQET Lagrangian employing $h_n$ is  
\be
\label{LObHQET}
\mc{L}_{\mr{bHQET}}^{(0)} = \bar{h}_n v\cdot iD \nn h_n, 
\ee
where we denoted the csoft derivative as $iD^{\mu}$, dropping the subscript for simplicity. 

Since the standard HQET Lagrangian, 
\be
\label{LOHQET} 
\mc{L}_{\mr{HQET}}^{(0)} = \bar{h}_v v\cdot iD  h_v,
\ee
has a boost invariance, it can be successfully applied to describe csoft interactions of the boosted heavy quark~\cite{Fleming:2007qr,Neubert:2007je}. 
Actually we find that leading results using bHQET are the standard results of HQET. For example, the field strength renormalization and the residue for $h_n$ at one loop are the same ones for $h_v$, 
\be
\label{stre}
Z_h = 1 + \frac{\as C_F}{2\pi} \frac{1}{\eps_{\mr{UV}}},~~~R_h = 1 - \frac{\as C_F}{2\pi} \frac{1}{\eps_{\mr{IR}}}.
\ee
So we might interpret bHQET as describing the boosted heavy quark interactions using the dominant portion ($h_n$) of the full spinor ($h_v$). The relation between $h_n$ and $h_v$ has been concretely considered in Appendix~\ref{bHQET} .

Employing bHQET introduced in Eq.~\eqref{LObHQET}, we can describe the heavy quark FFJ in the large $z$ limit. For this, we first integrate out collinear interactions and obtain the iHQJF, whose one-loop result was shown in Eq.~\eqref{IHQJF}. Then the heavy quark FFJ in Eq.~\eqref{defHQFFJ} is matched onto the following csoft function:
\bea
\label{csoftS}
S_{J/Q} (z;E_JR',m,\mu) &=& \sum_{X_{cs}} \frac{1}{2N_c} \mr{Tr} \frac{v_+}{2} \langle 0 |~ \delta\bigl((1-z)p_J^+ - i\partial_+\bigr) Y_{\n}^{cs\dg} h_n ~|J(p_J^+,\blpu{p}_J=0)X_{\notin J}^{cs}\rangle \nnb \\
&&\times \langle J(p_J^+,\blpu{p}_J=0) X_{\notin J}^{cs}|~ \bar{h}_n Y_{\n}^{cs} \nn ~| 0 \rangle,
\eea 
where $i\partial_+$ in the argument of delta function returns the csoft momentum of the parton that is not involved in the jet $J$. In matching from Eq.~\eqref{defHQFFJ}, Eq.~\eqref{Qmatch} has been applied, and $W_n$ has been replaced with $Y_{\n}^{cs}$. The dominant velocity component $v_+$ can be  approximated as $p_J^+/m$.

Finally, for the heavy quark FFJ in the large $z$ region, we have the following factorization: 
\be
\label{factFFJ}
D_{J/Q} (z\to 1;E_JR',m,\mu) = \mc{J}_Q (E_JR',m,\mu) S_{J/Q} (z;E_JR',m,\mu).
\ee
As we will see from the next-to-leading order (NLO) result of $S_{J/Q}$ in the next section, the typical csoft scale needed to minimize logarithms in $S_{J/Q}$ is given by $\mu_{cs} \sim (1-z)\sqrt{(E_JR')^2+m^2}$. So, through renormalization group (RG) evolution of the csoft function from the csoft scale to the collinear scale $\mu_c \sim \sqrt{(E_JR')^2+m^2}$, we will be able to resum large logarithms of $1-z$ near threshold.

If we consider the next-to-next-to-leading order in $\as$, we may have effects from the csoft gluons that are decoupled from the collinear gluons that form the heavy quark jet. Since the collinear gluon radiates in the limited phase space constrained by the jet boundary, the decoupled csoft gluons would give additional contributions to the csoft function in Eq.~\eqref{csoftS} and generate large nonglobal logarithms~\cite{Dasgupta:2001sh,Banfi:2002hw}. Resumming the nonglobal logarithms related to the heavy quark jet is beyond the scope of this paper,\footnote{
For a recent study of the nonglobal logarithms on the heavy quark, we refer to Ref.~\cite{Balsiger:2020ogy}.
} and here we will focus on the resummation of the `global' large logarithms of $1-z$ using the csoft function defined in Eq.~\eqref{csoftS}.

\section{NLO results of the heavy quark FFJ in the large $z$ limit and the resummation of large logarithms} 
\label{NLOFFJ}

\subsection{NLO calculation for the csoft function $S_{J/Q}$}

We first show that the tree level result of the csoft function, $S_{J/Q}^{(0)}$, has been normalized to $\delta(1-z)$. At tree level, Eq.~\eqref{csoftS} becomes 
\be
\label{csoftS0}
S_{J/Q}^{(0)} (z) = \frac{1}{2N_c} \sum_{s} \mr{Tr} \frac{v_+}{2} \langle 0 |~ \delta\bigl((1-z)p_J^+ \bigr)  h_n ~|Q_s(p_J^+)\rangle 
\langle Q_s(p_J^+) |~ \bar{h}_n  \nn ~| 0 \rangle,
\ee 
where $v_+=p_J^+/m$ and $Q_s$ denotes the heavy quark with spin $s$. Since the spin sum for the collinear (heavy) quark in $\mr{SCET_M}$ is given by
\be
\sum_s \xi_n |Q_s (p_+) \rangle  \langle Q_s (p_+) | \bar{\xi}_n = p_+ \nnn, 
\ee
using Eq.~\eqref{Qmatch}, we obtain the result for the heavy quark in bHQET 
\be
\label{spinsum} 
\sum_s h_n |Q_s (p_+) \rangle  \langle Q_s (p_+) | \bar{h}_n = 2m \nnn = m \fms{n}. 
\ee
Inserting Eq.~\eqref{spinsum} into Eq.~\eqref{csoftS0}, we obtain LO result, 
\be
S_{J/Q}^{(0)} (z) = \frac{1}{2}\cdot \frac{v_+}{2} \frac{\delta(1-z)}{p_J^+}~\mr{Tr}~m\fms{n}\nn =\delta(1-z). 
\ee  

\begin{figure}[t]
\begin{center}
\includegraphics[height=3.1cm]{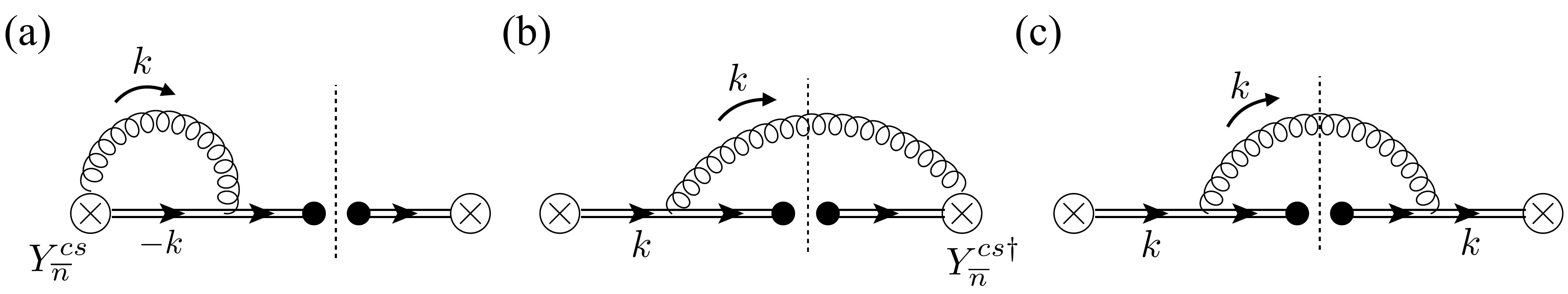}
\end{center}
\vspace{-0.8cm}
\caption{\label{fig2}\baselineskip 3.5ex
Feynman diagrams for the one-loop calculation of the csoft function $S_{J/Q}$. Double lines denote the bHQET field $h_n$, and $k$ is a csoft momentum. Virtual diagram (a) and real radiation diagram (b) have hermitian conjugate mirror diagrams. }
\end{figure}

We show the Feynman diagrams needed for the one-loop calculation of the csoft function $S_{J/Q}$ in Fig.~\ref{fig2}.
The virtual contribution shown in Fig.~\ref{fig2}-(a) is 
\be
\label{csvir1}
M_a = \mc{M}_a\cdot\delta(1-z) = ig_s^2 C_F \mu_{\overline{MS}}^{2\eps} \int\frac{d^D k}{(2\pi)^D} 
\frac{v_+}{(k^2+i\veps)(-v\cdot k +i\veps)k_+}\frac{\delta(1-z)}{p_J^+}.
\ee
Since we are working in a frame where $\blpu{p}_J=0$, we have $v_{\perp}=0$, and $v\cdot k$ is given by   
\be
v\cdot k = \frac{1}{2}\bigl(v_+ k_- + \frac{k_+}{v_+}\bigr) = \frac{1}{2} \bigl(\frac{p_J^+}{m}k_- + \frac{m}{p_J^+}k_+\bigr). 
\ee
After integrating over $k_-$ in Eq.~\eqref{csvir1}, we obtain
\bea
\mc{M}_a &=& -\frac{\as C_F}{2\pi} \frac{(\mu^2 e^{\gamma_E})^{\eps}}{\Gamma(1-\eps)} \int^{\infty}_0 \frac{dk_+}{k_+}
\int^{\infty}_0 d\blp{k}^2 \frac{(\blp{k}^2)^{-\eps}}{\blp{k}^2+m^2 k_+^2/p_J^{+2}} \nnb \\
\label{csvir2}
&=& - \frac{\as C_F}{2\pi} e^{\gamma_E} \Gamma(\eps) \Bigl(\frac{\mu^2}{m^2}\Bigr)^{\eps} \int^{\infty}_0 dx x^{-1-2\eps},
\eea
where $x\equiv k_+/p_J^+$, and the $\Gamma(\eps)$ regularizes the UV divergence. The 
IR pole for $x=0$ present in the remaining $x$ integral is problematic, since it cannot be regularized as $1/\eps_{\mr{IR}}$ because it overlaps with the UV divergence. However, it will cancel when combined with the real radiation contribution, as we will see. 

Diagrams (b) and (c) in Fig.~\ref{fig2} are contributions from real radiations, each of which can be separated into `in-jet' and `out-jet' contributions.  
For the in-jet contribution, the csoft gluon satisfies the jet criterion, $\theta<R'$, from Eq.~\eqref{jetmerg}. On the other hand, the radiation from the out-jet does not satisfy this constraint. The phase space constraints for the jet criterion were introduced in Eqs.~\eqref{pconq} and \eqref{pconp}, depending on our choice of frame. When a csoft gluon radiates, both can be simplified to 
\be
\blp{k}^2 < k_+^2 \tan^2 \frac{R'}{2} \sim k_+^2 \frac{R'^2}{4}.
\ee

The contribution from diagram (b) is 
\be
\label{csrea1}
M_b = \frac{\as C_F}{2\pi} \frac{(\mu^2 e^{\gamma_E})^{\eps}}{\Gamma(1-\eps)} \int^{\infty}_0 dk_+
\frac{p_J^+}{k_+} \delta \bigl((1-z)p_J^+-\Theta(\theta - R')k_+\bigr)
\int^{\infty}_0  \frac{d\blp{k}^2(\blp{k}^2)^{-\eps}}{\blp{k}^2+m^2 k_+^2/p_J^{+2}},
\ee
where $\Theta$ is the step function that returns 1 for the out-jet contribution. The in-jet contribution is proportional to $\delta(1-z)$. We can thus separate the contributions from  $M_b$ into  
\be
\label{Mbs}
M_b(z) = \mc{M}_b^{\mr{in}}\cdot \delta(1-z) + M_b^{\mr{out}} (z). 
\ee
Here $\mc{M}_b^{\mr{in}}$ is 
\be
\label{Mbin} 
\mc{M}_b^{\mr{in}}=\frac{\as C_F}{2\pi} \frac{(\mu^2 e^{\gamma_E})^{\eps}}{\Gamma(1-\eps)} \int^{\infty}_0 \frac{dx}{x}
\int^{x^2 E_J^2 R'^2}_0  \frac{d\blp{k}^2(\blp{k}^2)^{-\eps}}{\blp{k}^2+x^2 m^2},
\ee
where $x = k_+/p_J^+$ and $xE_JR' = k_+ R'/2$.  
The out-jet contribution $M_b^{\mr{out}}$ can be extracted from Eq.~\eqref{csrea1},
\be
\label{Mbout}
M_b^{\mr{out}}(z) = \frac{\as C_F}{2\pi} \frac{(\mu^2 e^{\gamma_E})^{\eps}}{\Gamma(1-\eps)} \frac{1}{1-z}
\int^{\infty}_{(1-z)^2 E_J^2 R'^2} \frac{d\blp{k}^2(\blp{k}^2)^{-\eps}}{\blp{k}^2+(1-z)^2 m^2}, 
\ee
where the large component of the csoft momentum, $k_+$, is given by $(1-z)p_J^+$. Eq.~\eqref{Mbout} has an IR divergence as $z$ goes to 1. In order to isolate the pole, we can employ the plus distribution 
\be
\label{Mbouts} 
M_b^{\mr{out}}(z) = \Bigl[\int^1_0 dz' M_b^{\mr{out}}(z') \Bigr]\delta(1-z) + \Bigl[M_b^{\mr{out}}(z)\Bigr]_+. 
\ee
The term with an integral can be written as 
\be
\label{Mboutint} 
\mc{M}_b^{\mr{out}} \equiv \int^1_0 dz M_b^{\mr{out}}(z) = \frac{\as C_F}{2\pi} \frac{(\mu^2 e^{\gamma_E})^{\eps}}{\Gamma(1-\eps)} \int^1_0 \frac{dx}{x} \int^{\infty}_{x^2 E_J^2 R'^2} \frac{d\blp{k}^2(\blp{k}^2)^{-\eps}}{\blp{k}^2+x^2 m^2},
\ee
where we have replaced $1-z$ with $x$. Eq.~\eqref{Mboutint} can be combined with Eq.~\eqref{Mbin}, since both are multiplied by the $\delta(1-z)$, and can be reorganized to be 
\bea
\mc{M}_b^{\mr{in}}+\mc{M}_b^{\mr{out}} &=& \frac{\as C_F}{2\pi} \frac{(\mu^2 e^{\gamma_E})^{\eps}}{\Gamma(1-\eps)} 
\Biggl\{\int^1_0 \frac{dx}{x} \int^{\infty}_{0} \frac{d\blp{k}^2(\blp{k}^2)^{-\eps}}{\blp{k}^2+x^2 m^2}
+\int^{\infty}_1 \frac{dx}{x} \int^{x^2E_J^2 R'^2}_{0} \frac{d\blp{k}^2(\blp{k}^2)^{-\eps}}{\blp{k}^2+x^2 m^2}\Biggr\} \nnb \\
\label{reorgMb}
&\equiv& \mc{M}_b^{A} + \mc{M}_b^{B}, 
\eea
where $\mc{M}_b^{A}~(\mc{M}_b^{B})$ corresponds to the first (second) term on the right-hand side of the first equality. 

Note that $\mc{M}_b^{A}$ has the same problematic pole (for $x=0$) that couples to the UV divergence that was in Eq.~\eqref{csvir2}. It cancels in the sum $\mc{M}_a+\mc{M}_b^{A}$, resulting in
\bea 
\mc{M}_a+\mc{M}_b^{A} &=& - \frac{\as C_F}{2\pi} e^{\gamma_E} \Gamma(\eps) \Bigl(\frac{\mu^2}{m^2}\Bigr)^{\eps} \int^{\infty}_1 dx x^{-1-2\eps} \nnb \\
\label{MaMbA}
&=&-\frac{\as C_F}{2\pi} \Bigl[\frac{1}{2\eps^2}+\frac{1}{2\eps}\ln\frac{\mu^2}{m^2} + \frac{1}{4}\ln^2 \frac{\mu^2}{m^2}+\frac{\pi^2}{24}\Bigr]. 
\eea 
Here the $1/\eps$ poles are all UV divergences. The calculation of remaining parts of $M_b$ is straightforward, and they only have UV divergences. We obtain $\mc{M}_b^B$ in Eq.~\eqref{reorgMb},
\be
\mc{M}_b^B = \frac{\as C_F}{2\pi} \Bigl[\frac{1}{2} \ln\frac{1+b}{b}\bigl(\frac{1}{\eps}+\ln\frac{\mu^2}{E_J^2R'^2}\bigr)+\frac{\pi^2}{12}+\frac{1}{4}\ln^2 b+\frac{1}{2} \mr{Li}_2(-b)\Bigr],
\ee
where $b=m^2/(E_JR')^2$. Finally, $[M_b^{\mr{out}}(z)]_+$ in Eq.~\eqref{Mbouts} is 
\be
\Bigl[M_b^{\mr{out}}(z)\Bigr]_+ = \frac{\as C_F}{2\pi} \Bigl[\frac{1}{1-z}\Bigl(\frac{1}{\eps}+\ln\frac{\mu^2}{(E_J^2R'^2+m^2)(1-z)^2}\Bigr)\Bigr]_+.
\ee

Moving on to the contribution of the diagram (c) in Fig.~\ref{fig2}, which is 
\bea
\label{csrealc}
M_c &=& -\frac{\as C_F}{\pi} \frac{(\mu^2 e^{\gamma_E})^{\eps}}{\Gamma(1-\eps)} \int^{\infty}_0 dk_+
\frac{k_+}{p_J^+} \delta \bigl((1-z)p_J^+-\Theta(\theta - R')k_+\bigr) \nnb \\
&&\times~ m^2\int^{\infty}_0  \frac{d\blp{k}^2(\blp{k}^2)^{-\eps}}{(\blp{k}^2+m^2 k_+^2/p_J^{+2})^2}.
\eea
Again, $M_c$ can be divided into in-jet and out-jet contributions, 
\be \label{Mcs}
M_c(z) = \mc{M}_c^{\mr{in}}\cdot \delta(1-z) + M_c^{\mr{out}} (z) 
= \bigl[\mc{M}_c^{\mr{in}}+\mc{M}_c^{\mr{out}}\bigr]\ \delta(1-z)+\bigl[M_c^{\mr{out}}(z)\bigr]_+,
\ee
where $\mc{M}_c^{\mr{out}} = \int^1_0 dz M_c^{\mr{out}}(z)$. 
After a brief computation we can again reorganize $\mc{M}_c^{\mr{in}}+\mc{M}_c^{\mr{out}}$, giving
\bea 
\label{Mcdel}
\mc{M}_c^{\mr{in}}+\mc{M}_c^{\mr{out}} &=& - \frac{\as C_F}{\pi} \frac{(\mu^2 e^{\gamma_E})^{\eps}}{\Gamma(1-\eps)} m^2  
\Biggl[\int^1_0 dx x \int^{\infty}_0  \frac{d\blp{k}^2(\blp{k}^2)^{-\eps}}{(\blp{k}^2+x^2 m^2)^2} \nnb \\
&&+\int^{\infty}_1 dx x \int^{x^2E_J^2R'^2}_0  \frac{d\blp{k}^2(\blp{k}^2)^{-\eps}}{(\blp{k}^2+x^2 m^2)^2}\Biggr], 
\eea
where the first term in the square brackets has an IR divergence as $x\to 0$ and the second has a UV divergence as $x\to \infty$. 
The IR divergence is removed by cancelling against the residue for the external heavy quark field. Including the one-loop results of the field strength renormalization and the residue for $h_n$ in Eq.~\eqref{Mcdel}, we obtain an IR finite result,
\be
\label{McZRdel}
\mc{M}_c^{\mr{in}}+\mc{M}_c^{\mr{out}} +Z_h^{(1)} + R_h^{(1)}  =
\frac{\as C_F}{2\pi} \Bigl[\frac{b}{1+b}\Bigl(\frac{1}{\eps}+\ln\frac{\mu^2}{E_J^2R'^2+m^2}\Bigr) - \frac{1}{1+b} \ln\frac{1+b}{b}\Bigr],
\ee 
where the renormalization $Z_h$ and the residue $R_h$ are given in Eq.~\eqref{stre}. $\bigl[M_c^{\mr{out}}(z)\bigr]_+$ in Eq.~\eqref{Mcs} is straightforwardly calculated to be 
\be
\label{Mcoutp}
\bigl[M_c^{\mr{out}}(z)\bigr]_+ = -\frac{\as C_F}{2\pi} \frac{2b}{1+b} \frac{1}{(1-z)_+}.    
\ee

We have now computed all the ingredients for one-loop calculation of the csoft function $S_{J/Q}$. The bare one-loop result is  
\bea 
S_{J/Q}^{(1)} (z) &=& 2(M_a + M_b) + M_c + (Z_h^{(1)}+R_h^{(1)})\delta(1-z) \nnb \\
&=& \bigl[2(\mc{M}_a+\mc{M}_b^{\mr{in}}+\mc{M}_b^{\mr{out}})+\mc{M}_c^{\mr{in}}+\mc{M}_c^{\mr{out}} +Z_h^{(1)} + R_h^{(1)}\bigr]\delta(1-z) \nnb \\
&&+ 2\bigl[M_b^{\mr{out}}(z)\bigr]_+ + \bigl[M_c^{\mr{out}}(z)\bigr]_+ \nnb \\
&=& \frac{\as C_F}{2\pi} \Biggl\{\delta(1-z) \Bigl[-\frac{1}{\eps^2}-\frac{1}{\eps} \ln \frac{\mu^2}{E_J^2 R'^2+m^2}
+\frac{b}{1+b}\bigl(\frac{1}{\eps}+\ln \frac{\mu^2}{E_J^2 R'^2+m^2}\bigr) \nnb \\
&&-\frac{1}{2} \ln^2 \frac{\mu^2}{E_J^2 R'^2+m^2}-\frac{1}{1+b} \ln (1+b) +\frac{1}{2}\ln^2(1+b)+\frac{\pi^2}{12}+\mr{Li}_2 (-b)\Bigr] \nnb \\
\label{SJQbo}
&& \Bigl[\frac{2}{1-z}\bigl(\frac{1}{\eps}+\ln\frac{\mu^2}{(E_J^2 R'^2+m^2)(1-z)^2}-\frac{b}{1+b}\bigr)\Bigr]_+ \Biggr\}, 
\eea
where the  $1/\eps$ poles are all UV divergences. 
When we take the massless limit, $m\to 0$~(equivalent to $b\to 0)$, we find Eq.~\eqref{SJQbo} is finite and recovers the result of the csoft function for the jet with a light quark calculated in Ref.~\cite{Dai:2017dpc}. So, just like $\mc{J}_Q$ in Eq.~\eqref{IHQJF}, we can interpret $S_{J/Q}$ in Eq.~\eqref{SJQbo} as the resummed result containing all of the $\mc{O}(m/E_JR')$ contributions when we consider the limit $E_JR' \gg m$. 

\subsection{RG evolution of the heavy quark FFJ for resummation of large logarithms}

In Sec.~\ref{factorization} we discussed the factorization of the heavy quark FFJ in the large $z$ limit. As shown in Eq.~\eqref{factFFJ}, the FFJ can be factorized into the iHQJF $\mc{J}_Q$ and the csoft function $S_{J/Q}$. Their NLO renormalized results are respectively 
\bea
\mc{J}_Q (E_JR',m,\mu) &=& 1+ \frac{\as C_F}{2\pi} \Biggl[ \frac{3+b}{2(1+b)} \ln\frac{\mu^2}{B^2} +\frac{1}{2} \ln^2 \frac{\mu^2}{B^2}+ \frac{1}{1+b}\bigl(2+\ln(1+b)\bigr)   \nnb \\
\label{IJFnlo}
&&~~-\frac{1}{2} \ln^2 (1+b) + f(b)+g(b)-\mr{Li}_2 (-b) + 2 -\frac{\pi^2}{12} \Biggr],
\eea
\bea
S_{J/Q} (z;E_JR',m,\mu) &=& \delta(1-z) 
+\frac{\as C_F}{2\pi} \Biggl\{\delta(1-z) \Bigl[
\frac{b}{1+b}\ln \frac{\mu^2}{B^2}-\frac{1}{2} \ln^2 \frac{\mu^2}{B^2}+\frac{\pi^2}{12}
\nnb \\
&&-\frac{1}{1+b} \ln (1+b) +\frac{1}{2}\ln^2(1+b)+\mr{Li}_2 (-b)\Bigr] \nnb \\
\label{SJQnlo}
&&+\Bigl[\frac{2}{1-z}\bigl(\ln\frac{\mu^2}{B^2(1-z)^2}-\frac{b}{1+b}\bigr)\Bigr]_+ \Biggr\}, 
\eea
where $B^2\equiv E_J^2R'^2+m^2$, and $f(b)$ and $g(b)$ are defined in Eqs.~\eqref{fb} and \eqref{gb} respectively. 

In the limit $z\to 1$, it is useful to take the Mellin transform of $S_{J/Q}$ and express the result in the large $N$ limit. The result is 
\bea 
\tilde{S}(\bar{N};E_JR',m,\mu)  &=& \int^1_0 dz z^{-1+N} S_{J/Q} (z;E_JR',m,\mu)\Bigl|_{N\to\infty} \nnb \\
&=& 1+ \frac{\as C_F}{2\pi}\Bigl[\frac{b}{1+b} \ln \frac{\mu^2\bar{N}^2}{B^2}-\frac{1}{2}\ln^2\frac{\mu^2\bar{N}^2}{B^2}-\frac{\pi^2}{4} \nnb \\
\label{mSJQ}
&&-\frac{1}{1+b} \ln (1+b) +\frac{1}{2}\ln^2(1+b)+\mr{Li}_2 (-b)\Bigr],
\eea 
where $\bar{N} \equiv N e^{\gamma_E}$, and the scaling of large $N$ is comparable with $1/(1-z)$. So we easily see that the typical csoft scale needed to minimize large logarithms in $S_{J/Q}$ is given by $\mu_{cs} \sim B(1-z)$.  

The function $\mc{J}_Q$ and $\tilde{S}_{J/Q}$ satisfies the following RG equations
\be
\label{RGEJS}
\frac{d}{d\ln\mu} f = \gamma_{f}\cdot f,~~~f=\mc{J}_Q,~\tilde{S}_{J/Q},
\ee 
where, to next-to-leading logarithmic (NLL) accuracy, the anomalous dimensions are given by 
\bea
\label{Janom} 
\gamma_{\mc{J}} &=& \Gamma_C \ln\frac{\mu^2}{B^2} + \hat{\gamma}_{\mc{J}},  \\
\label{Sanom}
\gamma_{S} &=& - \Gamma_C \ln \frac{\mu^2\bar{N}^2}{B^2} +\hat{\gamma}_{S},
\eea
with $\Gamma_C$ being the cusp anomalous dimension~\cite{Korchemsky:1987wg,Korchemskaya:1992je}, which can be expanded as $\Gamma_{C} = \sum_{k=0} \Gamma_{k}(\as/4\pi)^{k+1}$.
To NLL accuracy, we use its first two coefficients, given by 
\be
\Gamma_{0} = 4C_F,~~~\Gamma_{1} = 4C_F \Bigl[\bigl(\frac{67}{9}-\frac{\pi^2}{3}\bigr) C_A - \frac{10}{9} n_f\Bigr].
\ee
$\hat{\gamma}_{\mc{J},S}$ in Eqs.~\eqref{Janom} and \eqref{Sanom} are the non-cusp parts of the anomalous dimensions. 
For NLL accuracy, we use the leading (one-loop) results,   
\be
\label{hatJS}
\hat{\gamma}_{\mc{J}}^{(0)}=\frac{\as C_F}{2\pi}\frac{3+b}{1+b},~~
\hat{\gamma}_{S}^{(0)}=\frac{\as C_F}{2\pi}\frac{2b}{1+b}.
\ee
Note that the heavy quark FFJ, $D_{J/Q}(z\to1,\mu)=\mc{J}_Q(\mu)S_{J/Q}(z,\mu)$, is scale-dependent and follows DGLAP evolution.
This can be seen by combining the leading results of $\gamma_{\mc{J}}$ and $\gamma_{S}$ (at one loop), which has the result  
\be
\label{lcanom}
\gamma_{\mc{J}}^{(0)}+\gamma_{S}^{(0)} = \frac{\as C_F}{\pi}\bigl(\frac{3}{2}-2\ln\bar{N}\bigr),
\ee
which is the moment of the anomalous dimension for DGLAP evolution in the large $N$ limit. 

Solving the RG equations in Eq.~\eqref{RGEJS}, we can relate the factorized functions at the factorization scale $\mu_f$ to the functions at their natural scales, $\mu_{c}$ and $\mu_{cs}$, which minimize the logarithmic terms in $\mc{J}_Q$ and $S_{J/Q}$ respectively. Here the typical collinear scale for $\mc{J}_Q$ is given by $\mu_c \sim B$ and the csoft scale for $S_{J/Q}$ is $\mu_{cs} \sim B(1-z)$. Finally, after taking the inverse Mellin transform on $\tilde{S}_{J/Q}$, we obtain the following RG evolution result of the heavy quark FFJ:
\bea
D_{J/Q} (z\to 1;E_JR',m,\mu_f) &=& \mc{J}_Q (E_JR',m,\mu_f) S_{J/Q} (z;E_JR',m,\mu_f)\nnb \\
&=& \exp[\mc{M}(\mu_f,\mu_c,\mu_{cs})]\mc{J}_Q (E_JR',m,\mu_c)(1-z)^{-1+\eta}\nnb \\
\label{RGFFJ}
&&\times\tilde{S}_{J/Q}\bigl[\ln\frac{\mu_{cs}^2}{B^2(1-z)^2}-2\partial_{\eta}\bigr]\frac{e^{-\gamma_E \eta}}{\Gamma(\eta)}.
\eea 
This result automatically resums large logarithms of $1-z$, and resums large logarithms of small $R$ if the factorization scale is given by $\mu_f \sim E_J$. 

In Eq.~\eqref{RGFFJ} the argument $L$ of $\tilde{S}_{J/Q}[L]$ represents the logarithmic term in Eq.~\eqref{mSJQ}. In performing the inverse Mellin transform, the $\ln\bar{N}$ in Eq.~\eqref{mSJQ} can be converted to $-\partial_{\eta}$~\cite{Neubert:2005nt,Becher:2006nr}. We then used the relation 
\be
\tilde{S}_{J/Q}\bigl[\ln\frac{\mu_{cs}^2}{B^2}-2\partial_{\eta}\bigr](1-z)^{-1+\eta}\frac{e^{-\gamma_E \eta}}{\Gamma(\eta)}
=(1-z)^{-1+\eta}\tilde{S}_{J/Q}\bigl[\ln\frac{\mu_{cs}^2}{B^2(1-z)^2}-2\partial_{\eta}\bigr]
\frac{e^{-\gamma_E \eta}}{\Gamma(\eta)}.
\ee
The NLL exponentiation factor $\mc{M}$ in Eq.~\eqref{RGFFJ} is  
\bea
\mc{M}(\mu_f,\mu_c,\mu_{cs}) &=& -2S_{\Gamma}(\mu_c,\mu_{cs})-\ln\frac{\mu_c^2}{B^2} a_{\Gamma}(\mu_c,\mu_{cs}) \nnb \\
\label{expFFJ}
&&-\frac{C_F}{\beta_0}\Bigl[\frac{3+b}{1+b}\ln\frac{\as(\mu_f)}{\as(\mu_c)}+\frac{2b}{1+b}\ln\frac{\as(\mu_f)}{\as(\mu_{cs})}\Bigr],
\eea
where $S_{\Gamma}$ and $a_{\Gamma}$ are 
\be
S_{\Gamma} (\mu_1,\mu_2) = \int^{\alpha_1}_{\alpha_2} \frac{d\as}{b(\as)} \Gamma_{C}(\as) \int^{\as}_{\alpha_1}
\frac{d\as'}{b(\as')},~~~a_{\Gamma}(\mu_1,\mu_2) = \int^{\alpha_1}_{\alpha_2} \frac{d\as}{b(\as)} \Gamma_C(\as).
\ee
Here $\alpha_{1,2} \equiv \as (\mu_{1,2})$, and $b(\as)=d\as/d\ln\mu=-2\as\sum_{k=0}\beta_k (\as/4\pi)^{k+1}$ is the QCD beta function with $\beta_0$ in Eq.~\eqref{expFFJ} the leading coefficient. Finally the evolution parameter $\eta$ in Eq.~\eqref{RGFFJ} is  $\eta = 2a_{\Gamma}(\mu_f,\mu_{cs})$, which is  positive  for $\mu_f >\mu_{cs}$. 

\begin{figure}[t]
\begin{center}
\includegraphics[height=5cm]{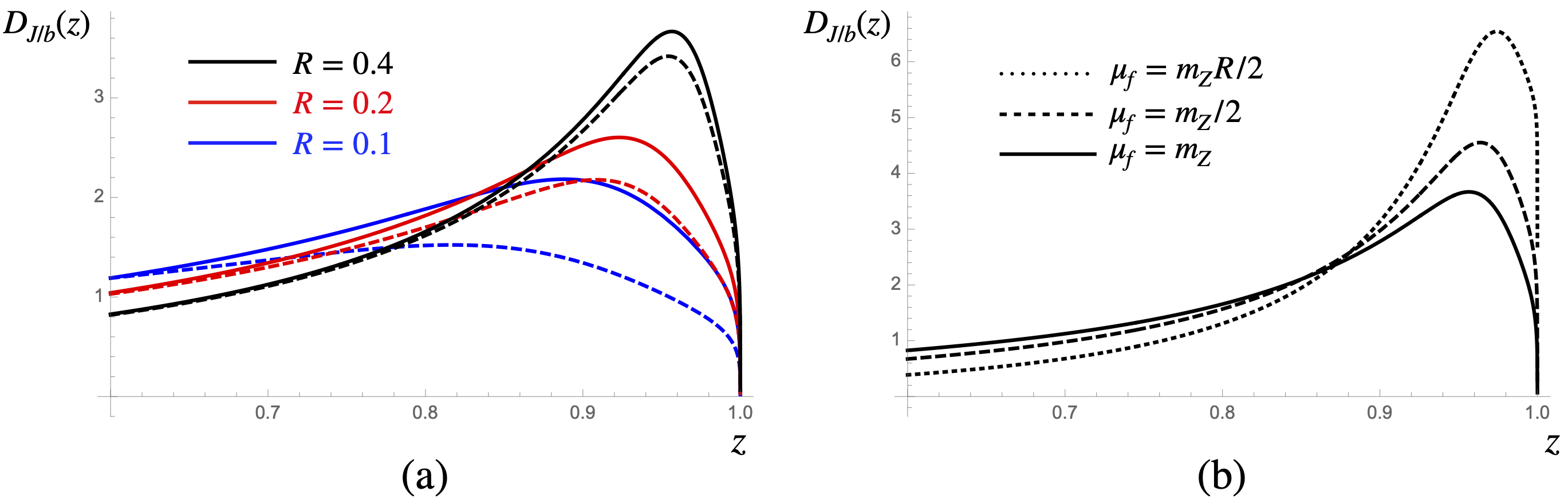}
\end{center}
\vspace{-0.8cm}
\caption{\label{hqffj1}\baselineskip 3.5ex
The $b$ quark FFJs near threshold for $e^+e^-$ annihilation at $Z$ pole. The maximal energy for the heavy quark jet and the factorization scale for the FFJs are respectively $E_J^{\max} = m_Z/2$ and $\mu_f = m_Z$.
In (a), the $b$ quark FFJs with $R=\{0.1,0.2,0.4\}$ are illustrated (solid lines) and compared with the results in the massless limit (dashed lines), while (b) shows the variance of the FFJs (with $R=0.4$) for different choices of the factorization scale. 
}
\end{figure}

In Fig.~\ref{hqffj1}~(a),  we show the resummed results of the $b$ quark FFJs near threshold for $e^+e^-$ annihilation at $Z$ pole for various choices of the jet radius $R$ (solid lines). 
The results have the accuracy of $\mr{NLL'}$, i.e., NLL in resummation plus NLO in the fixed order results.  
As $R$ increases, the FFJ becomes sharper and the distribution is populated in the region of larger $z$. We also show these results in the massless limit (dashed lines). The inclusion of the nonzero heavy quark mass significantly enhances the results compared to the massless limit. In Fig.~\ref{hqffj1}(b), we show the factorization scale $(\mu_f)$ dependence of the FFJ, which can cancel if we combine the FFJ with other factorized functions such as a hard function and possibly jet functions in other directions. 

\begin{figure}[h]
\begin{center}
\includegraphics[height=5cm]{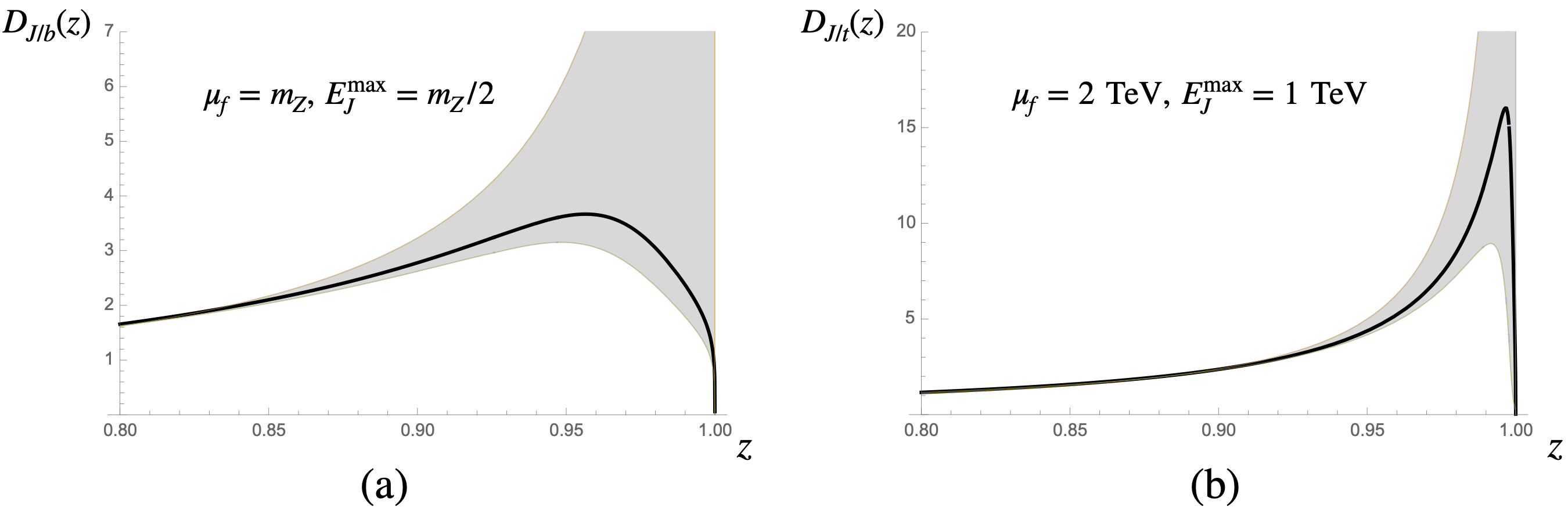}
\end{center}
\vspace{-0.8cm}
\caption{\label{hqffj3}\baselineskip 3.5ex
Uncertainties (gray bands) arising from choosing the jet scales between $B$ (the upper bound) and $B(1-z)$ (the lower bound) in (a) the $b$ quark FFJ and (b) the top quark FFJ when we employ the naive DGLAP evolution in Eq.~\eqref{naiveevo}. The solid lines denote the properly resummed results following Eq.~\eqref{RGFFJ} based on the factorization shown in Eq.~\eqref{factFFJ}.
}
\end{figure}

Unless we employ the factorization into the collinear and csoft parts shown in Eq.~\eqref{factFFJ}, we cannot properly resum the large logarithms of $1-z$. 
In this case, we may identify $\mu_c$ and $\mu_{cs}$ as a jet scale $\mu_J$, and consider the following evolution from the jet scale  to factorization scale, 
\be 
\label{naiveevo} 
D_{J/Q} (z\to 1;E_JR',m,\mu_f) = \exp[\tilde{\mc{M}}(\mu_f,\mu_J)] D_{J/Q} (z\to 1;E_JR',m,\mu_J).
\ee
Here the exponentiation factor $\tilde{\mc{M}}$ is given by Eq.~\eqref{expFFJ} with $\mu_c=\mu_{cs} =\mu_J$, i.e., 
$\tilde{\mc{M}}(\mu_f,\mu_J) = \mc{M}(\mu_f,\mu_J,\mu_{J})$. The heavy quark FFJ follows the naive DGLAP evolution with the anomalous dimension shown in Eq.~\eqref{lcanom}. 
We may select the jet scale $\mu_J$ for the FFJ to be some point between $B$ and $B(1-z)$. 
(However, any single choice of $\mu_J$ cannot minimize all the large logarithms in $D_{J/Q}$ of Eq.\eqref{naiveevo}.)
In Fig.~\ref{hqffj3}, we illustrate the large uncertainties in such cases, where the upper (lower) bounds of the gray bands denote the FFJ choosing the jet scale as $B~(B(1-z))$. Note that the scale choice of $\mu_J = B$ makes the FFJ blow up as $z$ goes to one in both Fig.~\ref{hqffj3}~(a) and (b).   

\begin{figure}[h]
\begin{center}
\includegraphics[height=4.7cm]{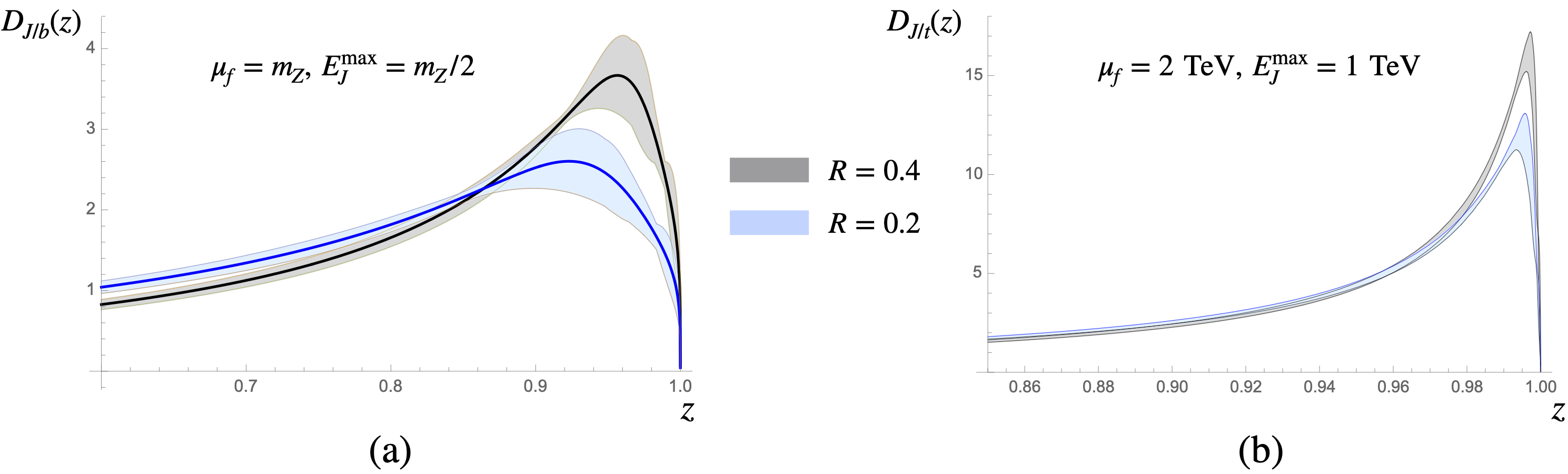}
\end{center}
\vspace{-0.8cm}
\caption{\label{hqffj2}\baselineskip 3.5ex
The collinear and the csoft scale variations of the resummed results for (a) the $b$ quark and (b) the top FFJs as described in the paper. The solid lines in  denote the results with the default scales given by $\mu_c = B$ and $\mu_{cs} = B(1-z)$. 
}
\end{figure}

In Fig.~\ref{hqffj2}, we have estimated errors due to variations of $\mu_{c}$ and $\mu_{cs}$ in the resummed results, Eq.~\eqref{RGFFJ}.  In obtaining the error bands in Fig.~\ref{hqffj2}, we varied $\mu_{c}$ and $\mu_{cs}$ from $\mu_{i(=c,cs)}^0/2$ to $2\mu_{i}^0$ respectively, where $\mu_i^0$ are the default scales given by $\mu_c^0=B$ and $\mu_{cs}^0=B(1-z)$.  
As shown in Fig.~\ref{hqffj2}~(b), the errors for the top FFJ with $\mu_f = 2~\mr{TeV}$ and $E_J^{\mr{max}}=1~\mr{TeV}$ are remarkably small, which may be due to the smallness of $\as$  at this very high energy scale.

\subsection{Heavy quark jet production near threshold from $e^+e^-$ annihilation}

When investigating heavy quark jets with small $R$ from $e^+e^-$ annihilation, the inclusive jet cross section can be written using the following factorization theorem:
\bea
\frac{d\sigma}{dx} &=& \sigma_0 \int^1_x \frac{dz}{z} \Bigl[H_Q(z;Q^2,\mu) D_{J/Q}\bigl(\frac{x}{z};E_JR,m,\mu\bigr)
+H_{\bar{Q}}(z;Q^2,\mu) D_{J/\bar{Q}}\bigl(\frac{x}{z};E_JR,m,\mu\bigr)\Bigr]\nnb \\
\label{fact1}
&=&2\sigma_0 \int^1_x \frac{dz}{z} H_Q(z;Q^2,\mu) D_{J/Q}\bigl(\frac{x}{z};E_JR,m,\mu\bigr),
\label{eq:xsection}
\eea
where $\sigma_0$ is the Born-level scattering cross section and $Q^2$ is the center of mass energy squared of the electron and the positron. In obtaining the second equality we ignored any charge asymmetry. 
$x$ is the heavy quark jet energy fraction, 
\be
x= \frac{2p_J\cdot q}{q^2}=\frac{2E_J}{Q}\sim\frac{p_J^+}{Q},
\ee
where $q^2=Q^2$, with $q$ being the momentum carried by the photon or $Z$ boson. The hard scattering contribution, $H_{Q}(z)$, is normalized to $\delta(1-z)$ at tree level, where $z$ is the energy fraction of the heavy quark. Since we will consider the  $Q\gg m$ limit, we can suppress the heavy quark mass dependence in $H_Q(z)$. 

Near threshold where $x$ is close to 1, the heavy quark FFJ can be factorized into $\mc{J}_Q$ and $S_{J/Q}$ as shown in Eq.~\eqref{factFFJ}. Additionally, $H_Q(z)$ can also be factorized as~\cite{Neubert:2007je,Fickinger:2016rfd}
\be
\label{factHJ}
H_Q(z;Q^2,\mu)=H(Q,\mu)J_{\n} (z;Q,\mu).
\ee
Here, the hard function $H(Q)$ includes virtual hard interactions with fluctuations of $Q^2$. The inclusive jet function $J_{\n}(z)$ describes (hard-)collinear interaction in the opposite direction of the heavy quark jet. The offshellness in $J_{\n}(z)$  is given by $p_X^2 \sim Q^2(1-z)$. To NLO in $\as$, $H(Q)$ and $J_{\n}(z)$ are 
\bea
\label{Hnlo}
H(Q,\mu) &=& 1+\frac{\as C_F}{2\pi} \Bigl(-3 \ln\frac{\mu^2}{Q^2}-\ln^2\frac{\mu^2}{Q^2}-8+\frac{7\pi^2}{6}\Bigr), \\
J_{\n} (z;Q,\mu) &=& \delta(1-z) +\frac{\as C_F}{2\pi}\Biggl\{\delta(1-z)\Bigl[\frac{3}{2}\ln\frac{\mu^2}{Q^2}
+\ln^2\frac{\mu^2}{Q^2}+\frac{7}{2}-\frac{\pi^2}{2}\Bigr] \nnb \\
\label{Jnlo}
&&\phantom{\delta(1-z) +\frac{\as C_F}{2\pi}\Biggl\{}
-\Bigl[\frac{1}{1-z}\bigl(2\ln\frac{\mu^2}{Q^2(1-z)}+\frac{3}{2}\bigr)\Bigr]_+\Biggr\}.
\eea

Finally, inserting Eqs.~\eqref{factFFJ} and \eqref{factHJ} into Eq.~\eqref{fact1}, we obtain the factorization theorem for the inclusive jet cross section near threshold 
\be
\label{factsigma}
\frac{1}{\sigma_0}\frac{d\sigma}{dx} =2 H(Q,\mu) \mc{J}_{Q} (E_JR',m,\mu) \int^1_x \frac{dz}{z} J_{\n} (z;Q,\mu) 
S_{J/Q}(z;E_JR',m,\mu). 
\ee
Based on this factorization, large logarithms of $R$ and $1-x$ can be simultaneously resummed through RG evolutions of the factorized functions. The resummed result is 
\bea
\frac{1}{\sigma_0}\frac{d\sigma}{dx} &=& 2 \exp[\mc{M}(\mu_h,\mu_{hc},\mu_c,\mu_{cs})] 
H(Q,\mu_h) \mc{J}_Q (E_JR',m,\mu_{c}) (1-x)^{-1+\eta} \nnb \\
\label{fact2}
&&\times\tilde{J}_{\n} \Bigl[\ln\frac{\mu_{hc}^2}{Q^2(1-x)}-\partial_{\eta}\Bigr] 
\tilde{S}_{J/Q}\bigl[\ln\frac{\mu_{cs}^2}{B^2(1-x)^2}-2\partial_{\eta}\bigr]\frac{e^{-\gamma_E \eta}}{\Gamma(\eta)},
\eea
where $\mu_h \sim Q$ and $\mu_{hc}\sim Q(1-x)^{1/2}$ are the scales that minimize the large logarithms in $H$ and $J_{\n}$, respectively. Note that the $\mu$-dependence shown in Eq.~\eqref{factsigma} is exactly cancelled in this resummed result. 
The evolution parameter $\eta$ is given by $\eta = 2a_{\Gamma}(\mu_{hc},\mu_{cs})$, which is a positive quantity. $\tilde{J}_{\n}$ is the Mellin transform of $J_{\n}$, with the NLO result 
\be
\tilde{J}_{\n}[L] = 1+\frac{\as C_F}{2\pi} \Bigl(\frac{3}{2}L+L^2+\frac{7}{2}-\frac{\pi^2}{3}\Bigr).
\ee
Finally, the exponentiation factor $\mc{M}$ in Eq.~\eqref{fact2} to NLL accuracy is
\bea
\mc{M}(\mu_h,\mu_{hc},\mu_c,\mu_{cs}) &=& 4 S_{\Gamma} (\mu_h,\mu_{hc}) - 2S_{\Gamma}(\mu_c,\mu_{cs})+2a_{\Gamma}(\mu_h,\mu_{hc}) \ln \frac{\mu_h^2}{Q^2} -a_{\Gamma}(\mu_c,\mu_{cs}) \ln \frac{\mu_c^2}{B^2}\nnb \\
\label{expsi}
&&-\frac{C_F}{\beta_0}\Bigl(3\ln\frac{\as(\mu_h)}{\as(\mu_{hc})}+\frac{3+b}{1+b}\ln\frac{\as(\mu_h)}{\as(\mu_{c})}
+\frac{2b}{1+b}\ln\frac{\as(\mu_h)}{\as(\mu_{cs})}\Bigr). 
\eea

\section{Heavy quark fragmentation to a groomed jet} 
\label{sec4}

As we have seen, the dominant contribution to the heavy quark FFJ comes from large $z$ region, and the (global) large logarithms of $1-z$ in this region can be successfully resummed through the factorization of the collinear and csoft contributions. But, if we use a grooming procedure on the heavy quark jet in order to investigate its substructure, the theoretical calculation of heavy quark fragmenting process can be severely modified because some csoft gluon radiations inside the jet are removed in the grooming process. In this and the following section, we consider a heavy quark fragmentation to a groomed jet focusing on large $z$ region. 
It will  be interesting to see how much the grooming changes the heavy quark FFJ in the large $z$ region. 

To be specific about the grooming procedure, we will consider soft drop~\cite{Larkoski:2014wba}.  
For $e^+e^-$ annihilation it can be implemented as follows:
\begin{enumerate}
\item Decluster the jet $j$ into two subjets $j_1$ and $j_2$ by undoing the last clustering process.
\item If the subjets satisfy the soft drop condition, 
\be\label{sdrop}
\frac{\mr{min}(E_{j_1},E_{j_2})}{E_{j_1}+E_{j_2}} > z_{\mr{cut}} \left( \frac{\theta_{12}}{R}\right)^{\beta},
\ee
then take $j$ to be the final soft-drop jet.
\item Otherwise redefine $j$ to be the subjet with the larger energy and go back to step~1.
\item If $j$ is a single particle or cannot be declustered further, we can either remove $j$ from the procedure (``the tagging mode'') or regard $j$ as the final soft-drop jet (``the grooming mode'').
\end{enumerate}
Here we start with the ungroomed jet $J$ with radius $R$ defined by one of the $\mr{k_T}$-type algorithms, and recluster the jet constituents into the subjets using the C/A algorithm. $\theta_{12}$ is the angle between the two subjets. 
In hadron collision,\footnote{\baselineskip 3.0 ex 
Since we apply the soft drop to a jet with small $R$, the groomed results do not modify hard interactions, only the fragmenting processes. So our results on the heavy quark fragmenting processes in sections \ref{sec4} and \ref{sec5} can be immediately applied to hadron collisions with replacement $E_JR \to p_T^JR$. 
} we use $p_T^j$ instead of $E_j$, and $\theta_{12}$ is replaced with $\Delta R = \sqrt{\Delta y^2+\Delta \phi^2}$.
If we choose the limit where the angular exponent $\beta$ vanishes, the procedure becomes equivalent to the mMDT procedure~\cite{Butterworth:2008iy}. So the soft drop can be considered as a generalized mMDT procedure.

In Eq.~\eqref{sdrop} the energy cut $\zc$ is usually set to be $\zc =0.1$, hence we will consider the limit $\zc \ll 1$. As we apply  soft drop to jets with small $R$, the parton or subjet that fails to satisfy the soft drop criterion in Eq.~\eqref{sdrop} can be regarded as csoft mode(s) in SCET. If we have a single parton after declustering, we will keep it as a final groomed jet, i.e., we adopt ``the grooming mode''. Doing so we do not exclude any event, and it is adequate for considering the fragmenting process to the jet. 

\subsection{Heavy quark jet fragmentation function to a groomed jet}

To begin, let us consider the probability for a groomed jet to take  energy fraction $z_G$ compared to  the ungroomed jet energy $E_J$. 
We will call this probability function the ``jet fragmentation function (JFF) to a groomed jet'', which also provides useful information on the energy loss when we groom a jet.  
The groomed jet energy differs from $E_J$ only when a radiated gluon inside the ungroomed jet has energy less than $\zc E_J$.
Hence the groomed jet energy fraction $z_G$ scales as $\sim 1-\zc$, and thus is close to 1. Therefore the fragmenting process to the groomed jet from the ungroomed jet is dominantly described by csoft interactions. 

In Refs.~\cite{Dai:2018ywt,Dai:2016hzf}, we formulated the differential cross section using the JFF for a given jet energy or $p_T$. Applying this formalism to the groomed jet from a heavy flavored jet, we describe the differential scattering cross section over $z_G$ and $E_J$ through the following factorization theorem: 
\be
\label{xseczg}
\frac{d\sigma}{dE_Jdz_G} = \int^1_{x=2E_J/Q}\frac{dz}{z} \frac{d\sigma (x/z;\mu)}{dE_i} D_{J_Q/i}(z;E_JR,\mu) \Phi_G (z_G),
\ee
where $Q$ is the center of mass energy of the electron and  positron for $e^+e^-$ annihilation. $D_{J_Q/i}$ is the fragmentation function to the heavy flavored jet~\cite{Dai:2018ywt}, and $\Phi_G$ is the heavy quark JFF to the groomed jet, which is determined by the csoft interactions. Note that $\Phi_G$ is a scale invariant quantity since the convolution of $d\sigma/dE_i$ and $D_{J_Q/i}$ is already scale invariant. 
If $E_J$ is not too close to $Q/2$, the jet energy fraction over the parton covers the full range of $z$, and $D_{J_Q/i}$ can be described by purely collinear interactions, while again $\Phi_G$ described by csoft interactions. Hence the factorization in Eq.~\eqref{xseczg} basically holds to all orders in $\as$. 

\begin{figure}[t]
\begin{center}
\includegraphics[height=6cm]{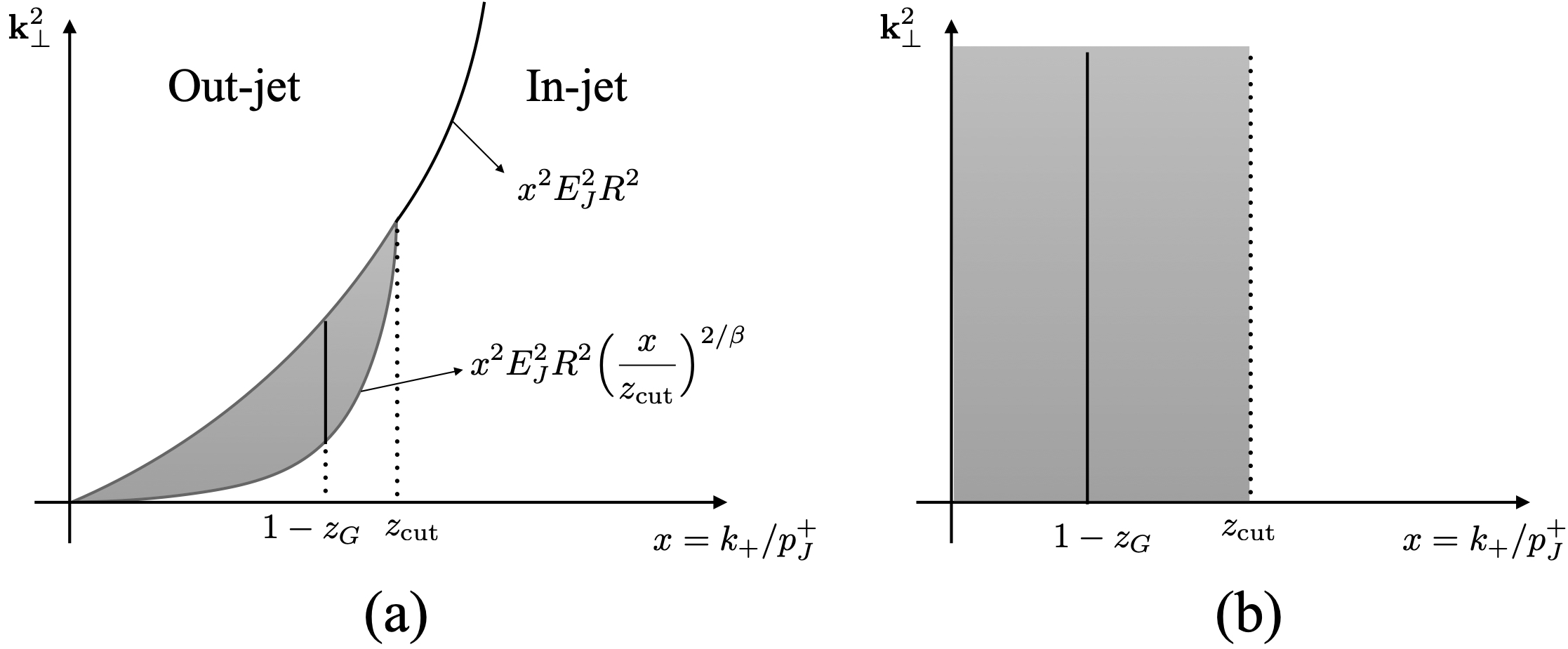}
\end{center}
\vspace{-0.8cm}
\caption{\label{fig3}\baselineskip 3.5ex
Diagram (a) is the phase space of the csoft gluon in the one-loop calculation of $\Phi_G$ for $E_JR \gtrsim m$. Diagram (b) denotes the phase space for the ultra-collinear soft (ucsoft) gluon radiation when we consider the limit $E_J R \gg m$ for $\Phi_G$ applying the soft drop with $\beta=0$. 
Both the shaded regions denote the groomed region, where the radiated gluon is dropped by the soft drop procedure. 
}
\end{figure}

Since the groomed jet involves an energetic heavy quark and the fragmenting process can be described entirely through csoft gluon radiations, we can apply bHQET to $\Phi_G$. Similar to Eq.~\eqref{csoftS}, $\Phi_G$ is defined as 
\bea
\label{PhiGdef}
\Phi_G (z_G;\zc, E_JR,m) &=& \sum_{X_{cs}} \frac{1}{2N_c} \mr{Tr} \frac{v_+}{2} \langle 0 |~ \delta\bigl((1-z_G)p_J^+ - \Theta_{\mr{in}}\cdot i\partial_+\bigr) Y_{\n}^{cs\dg} h_n ~|J_G X_{\notin J_G}^{cs}\rangle \nnb \\
&&\times \langle J_G(p_{J_G}^+,\blpu{p}_{J_G}=0) X_{\notin J_G}^{cs}|~ \bar{h}_n Y_{\n}^{cs} \nn ~| 0 \rangle,
\eea 
where $z_G = p_{J_G}^+/p_{J}^+$ and $J_G$ is the groomed jet. $\Theta_{\mr{in}}$ is the step function that equals one for gluon radiations to the in-jet region of $J_{Q}$ and zero otherwise. Hence $\Theta_{\mr{in}}\cdot i\partial_+$ returns the csoft momentum of the groomed gluon, i.e., the gluon to be dropped through grooming process. Eq.~\eqref{PhiGdef} is  normalized to $\delta(1-z_G)$ at LO in $\as$.   

In Fig.~\ref{fig3}(a), we illustrate the phase space for the radiated csoft gluon in the one-loop calculation of $\Phi_G$ applying the soft drop with the angular parameter $\beta \ge 0$. 
The shaded region denotes the groomed region, where $z_G$ can take  values other than $z_G=1$. In other regions, the contributions are  proportional to $\delta(1-z_G)$. Using the plus distribution on the real emission contribution to the groomed region, regularizing the IR divergence as $z_G$ goes to 1, the contribution can be written as 
\be
\label{gplus}
M_{R}^{gr} (z_G) = \delta(1-z_G) \left(\int_{gr} dx~ d\blp{k}^2 M_{R} (x,\blp{k}^2) \right) + \Bigl[\int_{gr} d\blp{k}^2 M_R (1-z_G,\blp{k}^2) \Bigr]_+,
\ee
where $M_R (x,\blp{k}^2)$ is the overall real emission distribution with $x=k_+/p_J^+$ and $\blp{k}^2$, with $k$ being the momentum of the radiated csoft gluon. In Eq.~\eqref{gplus}, the subscript `$gr$' in the integrals represents the phase space for the groomed region,  given by $0<x<\zc$ and $x^2 E_J^2 R^2 (x/\zc)^{2/\beta} < \blp{k}^2 < x^2 E_J^2 R^2$. In the groomed region $x$ can be also given by $1-z_G$. 

If we combine the term proportional to $\delta(1-z_G)$ in Eq.~\eqref{gplus} with the other real emission contributions outside of the groomed region, the integral of $M_{R} (x,\blp{k}^2) $ over $x$ and $\blp{k}^2$ covers the full phase space shown in Fig.~\ref{fig3}(a). Hence the net contribution proportional to $\delta(1-z_G)$ from the real emission must cancel the virtual contribution to $\Phi_G$ due to unitarity. 
Therefore the remaining nonvanishing contribution to $\Phi_G$ at one loop is given by the second term in Eq.~\eqref{gplus}, which is 
\bea
\Phi_G^{(1)} (z_G) &=& \Bigl[\int_{gr} d\blp{k}^2 M_R (1-z_G,\blp{k}^2) \Bigr]_+ \nnb \\
\label{phigol} 
&=& \frac{\as C_F}{\pi} \left[\frac{\Theta(z_G+\zc-1)}{1-z_G}\left(\ln\frac{1+b}{\bigl(\frac{1-z_G}{\zc}\bigr)^{2/\beta}+b} + \frac{b}{1+b} - \frac{b}{ \bigl(\frac{1-z_G}{\zc}\bigr)^{2/\beta}+b}\right)\right]_+, 
\eea 
where $b = m^2/(E_J^2R^2)$. This gives a nonzero value only when $z_G > 1-\zc$. As expected, the result has no scale dependence (except in $\as$). Note that  $\Phi_G$ has no large logarithm since $1-z_G \sim \zc$.
Also note that even if we consider the limit $1-z_G \ll \zc$, the growth in $\Phi_G$ is reasonable due to the presence of $b$. 

From Eq.~\eqref{phigol} we can also investigate the massless limit ($b=0$), where the result reads 
\be 
\label{phiGm0}
\Phi_G^{(1)} (z_G;m=0) =  \frac{\as C_F}{\pi} \frac{2}{\beta} \Bigl[\frac{\Theta(z_G+\zc-1)}{1-z_G} \ln \frac{\zc}{1-z_G}\Bigr]_+.
\ee
Note that this result only holds for $\beta >0$. If $\beta=0$, Eq.~\eqref{phiGm0} has an IR divergence, more specifically a collinear divergence, and hence a nonperturbative analysis is indispensable~\cite{Cal:2020flh,Marzani:2017mva}. 
However, in the massive case, Eq.~\eqref{phigol}, is IR safe for $\beta = 0$, with the result
\be
\label{phiGbeta0} 
\Phi_G^{(1)} (z_G;\beta=0) = \frac{\as C_F}{\pi} \Biggl[\frac{\Theta(z_G+\zc-1)}{1-z_G} \Bigl(\ln\frac{1+b}{b}-\frac{1}{1+b}\Bigr)\Biggr]_+.
\ee

If we consider the small $b$ limit $(E_JR \gg m)$,  large logarithm appears in Eq.~\eqref{phiGbeta0}. 
In order to resum the large logarithms through an additional factorization, we need to introduce a submode of the csoft mode.
We will call this mode the ``ultra-collinear soft (ucsoft) mode'', with the scaling behavior
\be 
\label{ucscale}
p_{ucs} = (p_{ucs}^+,p^{\perp}_{ucs},p_{ucs}^-) \sim E_J (1-z_G) \Bigl(1,\frac{m}{E_J},\frac{m^2}{E_J^2}\Bigr)
\sim E_J \zc \Bigl(1,\frac{m}{E_J},\frac{m^2}{E_J^2}\Bigr). 
\ee
Here the largest momentum component, $p_{ucs}^+$, is comparable with the csoft mode, i.e, $p_{ucs}^+ \sim p_{cs}^+$. But the 
offshellness, $p_{ucs}^2 \sim (1-z_G)^2 m^2 \sim \zc^2 m^2$, is much smaller than the csoft modes' offshellness in this limit. 

Using this to separate the csoft and the ucsoft modes in the limit $E_J R \gg m$, we factorize $\Phi_G$ with $\beta=0$ as 
\be 
\label{PhiGfac}
\Phi_G(z_G;\beta=0,E_JR \gg m) = \int^1_{z_G} \frac{dz}{z} S_G^{\Phi} (z;\zc,E_JR,\mu) U^{\Phi}_{G,c} (\frac{z_G}{z};\zc,m,\mu), 
\ee
where $S_G$ and $U_G$ are the csoft and ucsoft contributions respectively. As shown in Fig.~\ref{fig3}(b), the ucsoft mode scaling, Eq.~\eqref{ucscale}, does not recognize the jet boundary, since $R$ is much larger than $m/E_J$. Therefore, in calculating $U_G$ the upper limit of $\blp{k}^2$ can be taken to be infinity, resulting in an UV divergence. Since $S_G^{\Phi}$ is the matching coefficient between $\Phi_G$ and $U_{G,c}^{\Phi}$, and $\Phi_G$ is finite, $S_G^{\Phi}$ must have the same UV divergence as $U_G$, but with a relative minus sign.  To NLO in $\as$, the renormalized $S_G$ and $U^{\Phi}_{G,c}$ are 
\bea 
\label{SGPnlo} 
S_G^{\Phi} (z;\zc,E_JR,\mu) &=& \delta(1-z) - \frac{\as C_F}{\pi} \Biggl[\frac{\Theta(z+\zc-1)}{1-z} \ln\frac{\mu^2}{E_J^2R^2(1-z)^2}\Biggr]_+, \\
\label{UGnlo}
U^{\Phi}_{G,c} (z;\zc,m,\mu) &=& \delta(1-z) + \frac{\as C_F}{\pi} \Biggl[\frac{\Theta(z+\zc-1)}{1-z} \Bigl(\ln\frac{\mu^2}{m^2(1-z)^2}-1\Bigr)\Biggr]_+.
\eea 
Combining the one-loop results in Eqs.~\eqref{SGPnlo} and \eqref{UGnlo}, we recover Eq.~\eqref{phiGbeta0} in the limit $E_JR \gg m$.

\section{Heavy quark fragmentation function to a groomed jet in the large $z$ limit}
\label{sec5}

We now consider a heavy quark fragmenting to a groomed jet, which can be described by what we will call the heavy quark fragmentation function to a groomed jet (FFGJ), $D_{J_G/Q}(z)$. Since $\zc$ in the grooming procedure is given to be small, 
the effect of grooming can safely be ignored unless the energy fraction of the groomed jet over the heavy quark, $z$, is large. Therefore, in this section, we focus on the large $z$ behavior of the heavy quark FFGJ. 

As we see in Eq.~\eqref{factFFJ}, when we consider the heavy quark fragmenting to a jet in the large $z$ limit, the heavy quark FFJ can be factorized into iHQJF and $S_{J/Q}$. Instead of an ungroomed jet, if we consider a groomed jet, the grooming effects will modify the csoft function $S_{J/Q}$. 
So, introducing a new csoft function including grooming, we formulate the heavy quark FFGJ in the large $z$ region as 
\be \label{DJGQ}
D_{J_G/Q} (z;E_JR,m,\mu) = \mc{J}_Q (E_JR,m,\mu) S_{J_{G/Q}} (z;\zc,E_JR,m,\mu). 
\ee

Since the ungroomed csoft function, $S_{J/Q}$, describes the out-jet radiations, its one-loop contribution does not feel the effects of grooming. Therefore, to NLO in $\as$, the csoft function for the groomed jet can simply be given by  
\be
S_{J_G/Q} (z;\zc, E_JR,m,\mu) = \delta(1-z) + S_{J/Q}^{(1)} (z) + \Phi_{G}^{(1)} (z), 
\ee 
where the one-loop contributions, $S_{J/Q}^{(1)}$ and $\Phi_{G}^{(1)}$, have been obtained in Eqs.~\eqref{SJQnlo} and \eqref{phigol}, respectively. Since $\Phi_{G}^{(1)}$ returns a nonzero value only in the region $z>1-\zc$, $S_{J_G/Q}^{(1)}$ is equal to $S_{J/Q}^{(1)}$ in the region $z<1-\zc$. Thus the grooming effects modify the original heavy quark FFJ only in the region $z>1-\zc$. 

\begin{figure}[t]
\begin{center}
\includegraphics[height=6cm]{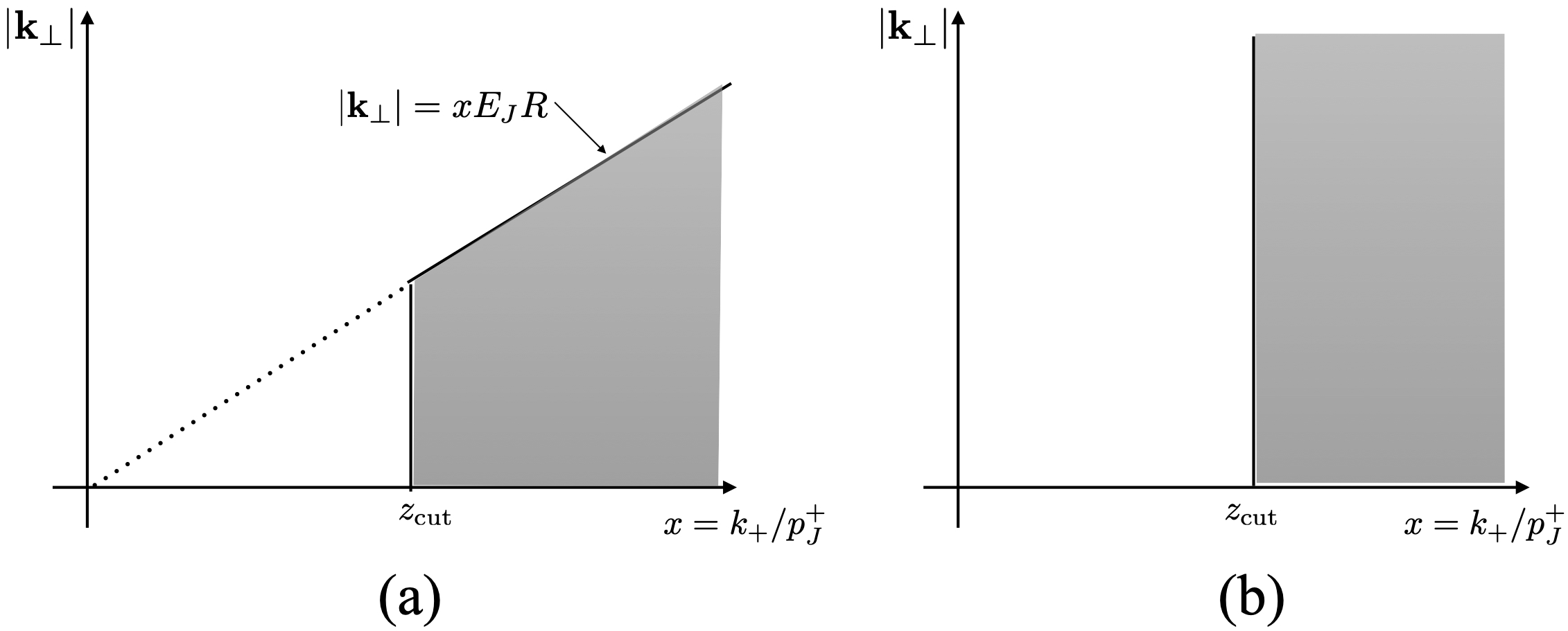}
\end{center}
\vspace{-0.8cm}
\caption{\label{fig4}\baselineskip 3.5ex
(a): Phase space for a radiated gluon in $S_{J_G/Q}$. Here the shaded region is the inner space of the groomed jet and the contribution is proportional to $\delta(1-z)$. The contributions with $z\neq 1$ comes from the outside of the shaded region. 
(b): The phase space for the ucsoft gluon scaling as $p_{ucs} \sim \zc E_J(1,m/E_J,(m/E_J)^2)$, which is decoupled from the csoft mode when we consider the limit $E_JR\gg m$. 
}
\end{figure}

For simplicity, we consider the grooming effects using the soft drop with $\beta = 0$. In this case the fragmenting process in the massless limit becomes IR (collinear) divergent. However, in the fragmenting process for a heavy-flavored jet, the heavy quark mass prevents the divergence. In Fig.~\ref{fig4}(a), the phase space for a radiated gluon in the function $S_{J_G/Q}$ is illustrated. Here the space inside of the groomed jet is denoted as the shaded region, in which the contribution is proportional to $\delta(1-z)$ at NLO in $\as$. In the outer region the gluon splits away. In the massless case, we have a collinear divergence from the region $x\in[0,\zc]$ at $|\blp{k}|=0$, which is regularized in the massive case by the heavy quark mass.    

\subsection{Factorization of the heavy quark FFGJ in the limit $1-z\ll \zc$}

Since the grooming affects the region of $z$ above $1-\zc$, we need to investigate this region more closely. 
The dependence of $\zc$ in this region gives rise to an additional large logarithm near the end point, i.e., $\ln [\zc/(1-z)]$. 
In order to handle the large logarithm, we need to refactorize $S_{J_G/Q}$ in the region. Since $1-z$ is much smaller than $\zc$ near the end point,  
a new softer mode can be decoupled from the csoft mode. We call it the ``collinear-ultrasoft (cusoft)'' mode, and it scales as 
$p_{cus}\sim (1-z) E_J(1,R,R^2)$, while the csoft mode in this region scales as $p_{cs} \sim \zc E_J(1,R,R^2)$. 

The csoft mode near the end point cannot radiate out of the groomed jet; the available phase space for the radiation is given by the shaded region in Fig.~\ref{fig4}(a). The one-loop contribution of the csoft mode is obtained by integrating over the shaded region, 
\be
\label{csgol}
S_G^{(1)} = \int^{\infty}_{\zc} dx \int^{x^2 E_J^2 R^2}_0 d\blp{k}^2 M_R(x,\blp{k}^2),
\ee
where $M_R(x,\blp{k}^2)$ is the real emission distribution in the csoft limit as described below Eq.~\eqref{gplus}. 
For the complete one-loop result, we need to include the virtual contribution. But, when we factorize $S_{J_G/Q}$ decoupling the cusoft mode from the csoft mode, we need to subtract the cusoft contributions in the calculation of the csoft part. This is the conventional matching procedure, i.e, the zero-bin subtraction. In this subtraction, the virtual contribution cancels since the virtual contributions for both the modes are the same. Therefore the one-loop result entirely comes from the integration of Eq.~\eqref{csgol}, and the renormalized NLO result for the csoft function $S_G$ is 
\bea
S_G (\zc, E_JR, m,\mu)  &=& 1+ \frac{\as C_F}{2\pi} \Bigl[\Bigl(\ln\frac{1+b}{b} - \frac{1}{1+b}\Bigr) \ln\frac{\mu^2}{\zc^2 E_J^2R^2} + \frac{\pi^2}{6} \nnb \\
\label{SGnlo}
&&\phantom{1+ \frac{\as C_F}{2\pi} \Bigl[}
+ \frac{1}{2} \ln^2 b + \mr{Li}_2 (-b) - \ln\frac{1+b}{b}\Bigr]. 
\eea 

When we consider the cusoft contribution in the factorization, we focus on the region near $x=0$ in Fig.~\ref{fig4}(a), where $x$ is equivalent to $1-z$. In this region, the split gluon from the jet can radiate either inside or outside the jet. Furthermore, the gluon cannot recognize $\zc$ since the cusoft mode is much softer than the csoft mode. Note that the cusoft mode describes the same situation as the case of the (partonic) heavy quark fragmentation function (HQFF). In the large $z$ region, the  HQFF can be additionally factorized into the virtual heavy quark function and the shape function~\cite{Neubert:2007je,Fickinger:2016rfd}. 
The cusoft gluon here can radiate into the full phase space like the soft gluon does in the HQFF. Therefore, the cusoft contribution to one loop should be the same as the shape function for the HQFF and the NLO result is thus 
\bea
U_{cus} (z;m,\mu) &=& \delta(1-z) + \frac{\as C_F}{2\pi} \Bigl\{\delta(1-z) \Bigl[\ln\frac{\mu^2}{m^2} - \frac{1}{2}\ln\frac{\mu^2}{m^2} - \frac{\pi^2}{12} \Bigr] \nnb \\
\label{shape} 
&&\phantom{\delta(1-z) + \frac{\as C_F}{2\pi} \Bigl\{}
+ \Bigl[\frac{2}{1-z} \Bigl(\ln\frac{\mu^2}{m^2(1-z)^2}-1\Bigr) \Bigl]_+\Bigr\}.   
\label{eq:Ucus}
\eea 
Here the large logarithms are minimized when $\mu \sim m(1-z)$. 

Since the cusoft function in Eq.~\eqref{shape} can be given independent of knowledge of the jet boundary (i.e., no $R$ dependence), we can express the scaling of the cusoft mode more precisely as 
\be 
\label{cusoft}
p_{cus} \sim (1-z)E_J \Bigl(1,\frac{m}{E_J}, \frac{m^2}{E_J^2}\Bigr).  
\ee
This will be important if we consider the limit $E_JR \gg m$. Finally, the factorization formula for $S_{J_G/Q}$ near the end point is 
\be 
\label{facgS1}
S_{J_G/Q} (z\to1;\zc, E_JR,m,\mu) =  S_G (\zc, E_JR, b,\mu) U_{cus} (z;m,\mu). 
\ee
Here the cusoft function is free of nonglobal logarithms. The dependence on the logarithms resides only in $S_{G}$. Since the contribution to $S_{G}$ comes from the region of $z$ below $1-\zc$, the nonglobal logarithm effects from the grooming is limited~\cite{Larkoski:2014wba,Dasgupta:2013ihk}. 

\begin{figure}[htb]
	\begin{center}
		\includegraphics[height=7cm]{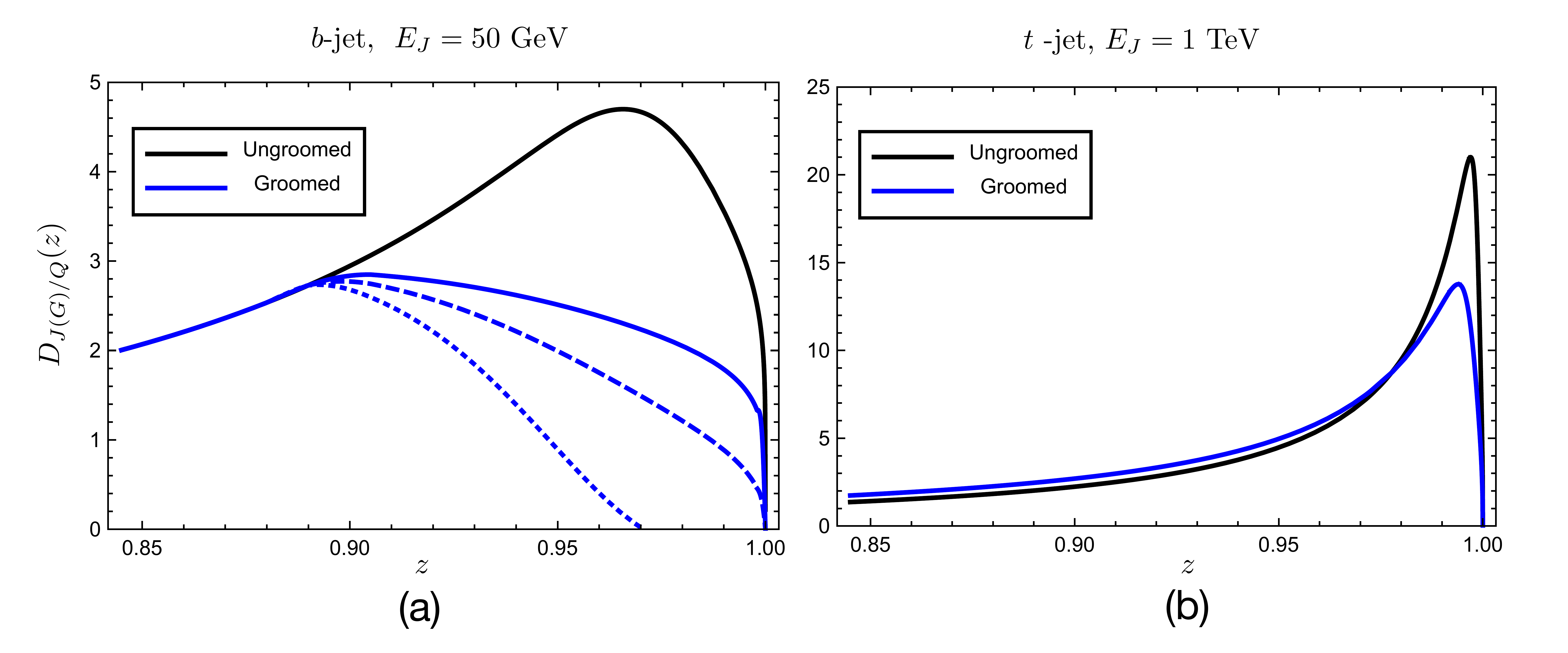}
	\end{center}
	\vspace{-0.8cm}
	\caption{\label{fig:FFJNoH} 
		FFJ and FFGJ for (a) bottom and (b) top quark jets near threshold, with no separation between jet scale and heavy quark mass, i.e., $E_J R \sim m_{b(t)}$.  The black curves are FFJs and the blue curves are FFGJs, with $z_{\rm cut}=0.1$, $R=0.4$, and $\mu_{min}=0.5$\,GeV for the both solid black and solid blue curves (Ref.~Appendix~\ref{sec:profile} for the definition $\mu_{min}$ in profile functions). In panel (a), the dashed curve corresponds to $\mu_{min}=0.4$\,GeV in the profile function for the ultrasoft scale $\mu_{us}$, and the dotted curve corresponds to $\mu_{min}=0.3$\,GeV. Note that for the top quark jet, the groomed FFJ is insensitive to the exact value of $\mu_{min}$. 
	}
\end{figure}

In Fig.~\ref{fig:FFJNoH}, with accuracy $\mr{NLL'}$, we show the comparison between FFJ and FFGJ for bottom and top quark jets, in the limit $1-z \ll z_{\rm cut}$ and with no separation of scale between $E_JR$ and heavy quark mass $m$, where $z_{\rm cut}=0.1$. The factorization scale of both the FFJ and FFGJ are set to be $\mu_f=E_J$. For the bottom quark jet, Fig.~\ref{fig:FFJNoH}(a), the grooming effect is large. However, the collinear-ultrasoft scale $\mu_{cus}=m_b(1-z) \ll 0.48\text{\,GeV}(=m_b \zc)$ in the limit $1-z \ll z_{\rm cut}$, so non-perturbative effects are significant. We have chosen different profile functions to test this (for a detailed description of profile functions used in this paper, please refer to Appendix \ref{sec:profile}; the purpose of the profile functions is to freeze the $z$-dependent scale to a minimum value $\mu_{min}$ as $z\to1$). In Fig.~\ref{fig:FFJNoH}(a), the solid, dashed, and dotted blue curves corresponds to $\mu_{min}=0.5$, $0.4$, and $0.3$\,GeV, which shows the sensitivity to non-perturbative effects. 

In Fig.~\ref{fig:FFJNoH}(a), we have interpolate the FFGJs with the FFJ in a tiny region $(z_{\rm cut}-0.05,z_{\rm cut}+0.05)$, so that the FFGJs smoothly transform to the FFJ in the region $z<z_{\rm cut}$. This is fine since the factorization Eq.~\eqref{facgS1} used here applies only to the limit $1-z \ll z_{\rm cut}$. For the top quark jet, Fig.~\ref{fig:FFJNoH}(b), the cusoft scale $m_{t} (1-z)$ is still perturbative in the limit $1-z \ll z_{\rm cut}$, as long as $z$ is not too close to $1$. Note that in panel (b) of Fig. \ref{fig:FFJNoH}, $E_JR \sim 2 m_t$, so $E_J R$ can still be reasonably treated as the same scale of $m_t$, which justifies the use of Eq.~(\ref{facgS1}).  The more interesting case is when we impose the scale hierarchy $E_JR \gg m$, as discussed in the next section.

\subsection{Refactorization of the csoft function $S_G$ when $E_JR \gg m$}

In Eq.~\eqref{PhiGfac}, when considering the limit $E_JR \gg m$, we factorized the JFF to a groomed jet, $\Phi_G$ into the csoft and ucsoft parts. Similarly for $S_G$ in Eq.~\eqref{facgS1}, we can decouple the ucsoft mode from the csoft mode, which scales as 
\be
p_{ucs} \sim \zc E_J \Bigl(1,\frac{m}{E_J},\frac{m^2}{E_J^2}\Bigr).  
\ee
Since we are considering the limit $1-z \ll \zc$, the ucsoft mode here does not contribute to the actual distribution of $z$. 
Instead $S_G$ in Eq.~\eqref{facgS1} can be refactorized into the csoft and the ucsoft parts
\be 
\label{factSG}
S_G(\zc,E_JR\gg m,\mu) = S_{G,0} (\zc,E_JR,\mu) U_{G,ucs} (\zc,m,\mu). 
\ee
Here $S_{G,0}$ is the csoft function with the mode $p_{cs} \sim \zc E_J (1,R,R^2)$, and $U_{G,ucs}$ is a newly introduced ucsoft function. 

Since the ucsoft gluon radiates over too narrow an angle to distinguish the jet boundary, the upper limit of the transverse momentum to the heavy quark, $|\blp{k}|$, can be taken to be infinity. The available phase space of the ucsoft gluon at one loop is shown by the shaded region in Fig.~\ref{fig4}(b). When we compute the ucsoft function to one loop, the virtual contribution vanishes.
Since the cusoft mode~$(p_{cus}^2 \sim (1-z)^2 m^2)$, which scales as Eq.~\eqref{cusoft}, can be regarded as a submode of the ucsoft mode ~$(p_{ucs}^2 \sim \zc^2 m^2)$, the same virtual contribution from the cusoft mode needs to be subtracted. Hence the complete one-loop result can be obtained from integration of the shaded region in Fig.~\ref{fig4}(b), and the NLO result is 
\be 
\label{UGcnlo}
U_{G,ucs}(\zc,m,\mu) = 1+ \frac{\as C_F}{2\pi} \Bigl(-\ln\frac{\mu^2}{\zc^2 m^2}+\frac{1}{2} \ln^2\frac{\mu^2}{\zc^2 m^2}+\frac{\pi^2}{12} \Bigr). 
\ee

The csoft function $S_{G,0}$ in Eq.~\eqref{factSG} is the matching coefficient between $U_{G,c}$ and $S_G$ in the limit $E_JR \gg m$. Hence, subtracting the one-loop result in Eq.~\eqref{UGcnlo} from the result of $S_G$ in Eq.~\eqref{SGnlo} and taking the limit as $m\to 0$, we can obtain the one-loop result of $S_{G,0}$,\footnote{\baselineskip 3.0 ex 
The result can also be obtained from the integration of the shaded region after subtracting the region shown in Fig.~\ref{fig4}(b) from the one shown in Fig.~\ref{fig4}(a).  
} 
\be 
S_{G,0} (\zc,E_JR,\mu) = 1 + \frac{\as C_F}{2\pi} \Bigl(-\frac{1}{2}\ln^2\frac{\mu^2}{\zc^2 E_J^2 R^2}+\frac{\pi^2}{12} \Bigr).
\label{eq:SG0}
\ee
The cusoft function, $U_{G,s}$, is unchanged in the limit $E_JR \gg m$. Therefore $S_{J_G/Q}$ in this limit is given by
\be 
\label{facgS2}
S_{J_G/Q} (z\to1;\zc, E_JR\gg m,\mu) =  S_{G,0} (\zc,E_JR,\mu) U_{G,ucs} (\zc,m,\mu)U_{cus} (z;m,\mu). 
\ee

\subsection{Resummed results of the heavy quark FFGJ}

We now turn to the resummed result of the heavy quark FFGJ, beginning with the assumption $E_J R\gtrsim m$. As shown in Eq.~(\ref{DJGQ}), the grooming modifies the soft function by the addition of a new csoft function in the factorized form.  We can write
\bea
D_{J_G/Q} (z\to 1;E_JR,m,\mu) &=& \mc{J}_Q (E_JR,m,\mu) S_{J_G/Q} (z\to1;E_JR,m,\mu_f) \\
&=& \mc{J}_Q (E_JR,m,\mu) S_{G}(\zc, E_JR, b, \mu) U_{cus} (z\to1; m, \mu). \nnb 
\eea 
The resummation of the iHQJF, $\mc{J}_Q$, was discussed previously. We have
\be
\mc{J}_Q(\mu_f) = \exp[\mc{M}_c (\mu_f,\mu_c)] \mc{J}_Q(\mu_c),
\ee
where 
\be
\mc{M}_c(\mu_f,\mu_c) = 2 S_\Gamma(\mu_f,\mu_c) + a_\Gamma(\mu_f,\mu_c) \ln\frac{\mu_f^2}{B^2} + a_{\hat\gamma_c}(\mu_f,\mu_c),
\ee
with 
\be
a_{\hat\gamma_c}(\mu_f,\mu_c) = -\frac{C_F}{\beta_0} \ln\frac{\alpha_s(\mu_f)}{\alpha_s(\mu_c)} \frac{3+b}{1+b}.
\ee

Turning to the resummation of the soft pieces, from Eq.~(\ref{SGnlo}), we obtain
\be
S_G(\mu_f) = \exp[\mc{M}_{cs} (\mu_f,\mu_{cs}) ]S_G(\mu_{cs}) ,
\ee
where
\be
\mc{M}_{cs} (\mu_f,\mu_{cs}) = a_{\hat \gamma_{cs}}(\mu_f,\mu_{cs}) = -\frac{2C_F}{\beta_0} \ln\frac{\alpha_s(\mu_f)}{\alpha_s(\mu_{cs})} \left[\ln\frac{1+b}b - \frac1{1+b} \right].
\ee
Since we are interested in the $z\to1$ limit, we take the Mellin transform of $U_{G,s}$, in the large $N$ limit, obtaining
\be
\tilde U_{cus}(\bar N;E_JR,m,\mu) = 1 +\frac{\alpha_S C_F}{2\pi} \left[\ln\frac{\mu^2\bar N^2}{m^2} - \frac12\ln^2\frac{\mu^2\bar N^2}{m^2} - \frac5{12} \pi^2\right].
\ee
This is resummed by
\be
\tilde U_{cus}(\bar N;\mu_f) = \exp[\mc{M}_{cus}(\mu_f,\mu_{cus}) ] \bar N^{-\eta_U} \tilde U_{cus} (\bar N, \mu_{cus}),
\ee
where
\be
\mc{M}_{cus}(\mu_f,\mu_{cus})  = -2S_\Gamma(\mu_f,\mu_{cus}) - \ln\frac{\mu_f^2}{m^2} a_\Gamma(\mu_f,\mu_{cus}) + a_{\hat\gamma_{cus}}(\mu_f,\mu_{cus}),
\ee
with 
\be
a_{\hat\gamma_{cus}}(\mu_f,\mu_{cus}) = -\frac{2C_F}{\beta_0} \ln\frac{\alpha_s(\mu_f)}{\alpha_s(\mu_{cus})}, 
\ee
and
\be
\eta_U = 2 a_\Gamma (\mu_f, \mu_{cus}).
\ee
Combining everything, we obtain
\bea
D_{J_G/Q} (z\to 1;E_JR,m,\mu_f) &=& \mc{J}_Q (E_JR,m,\mu_f) S_{J_G/Q} (z;E_JR,m,\mu_f) \\
&=& \exp[\mc{M}(\mu_f,\mu_c,\mu_{cs},\mu_{cus})]\mc{J}_Q (E_JR,m,\mu_c)(1-z)^{-1+\eta}\nnb \\
\label{RGFFJG1}
&&\times S_{G}(\zc, E_JR, b, \mu_{cs})\tilde{U}_{cus}\bigl[\ln\frac{\mu_{cus}^2}{m^2(1-z)^2}-2\partial_{\eta}\bigr]\frac{e^{-\gamma_E \eta}}{\Gamma(\eta)},\nnb
\eea 
where 
\bea
\mc{M}(\mu_f,\mu_c,\mu_{cs},\mu_{cus}) &=& -2 S_\Gamma(\mu_c,\mu_{cus}) + \ln\frac{m^2}{\mu_c^2} a_\Gamma(\mu_c,\mu_{cus}) + \ln\frac{m^2}{B^2} a_\Gamma(\mu_f,\mu_c) \\
&& -\frac{C_F}{\beta_0}\left[\frac{3+b}{1+b}\ln\frac{\alpha_s(\mu_f)}{\alpha_s(\mu_c)} +2\left(\ln\frac{1+b}{b} -\frac1{1+b}\right) \ln\frac{\alpha_s(\mu_f)}{\alpha_s(\mu_{cs})} +  2 \ln\frac{\alpha_s(\mu_f)}{\alpha_s(\mu_{cus})}\right],\nnb
\eea
and $\eta =\eta_U$.

We now turn to the region where $E_JR \gg m$, in which case we can further factorize  $S_G$, as shown in Eq.~(\ref{factSG}), with the addition of the new ucsoft function $U_{G,ucs}$.
This modifies the above to be\footnote{Here the heavy quark mass dependence in iHQJF $\mc{J}_Q$ can be safely ignored.}
\bea
\label{RGFFJG2}
D_{J_G/Q} (z\to 1;E_JR,m,\mu_f) &=& \mc{J}_Q (E_JR,m,\mu_f) S_{J_G/Q} (z;E_JR,m,\mu_f) \\
&=& \exp[\mc{M}(\mu_f,\mu_c,\mu_{cs},\mu_{ucs},\mu_{cus})]\mc{J}_Q (E_JR,m,\mu_c)S_{G,0}(\zc, E_JR, \mu_{cs})  \nnb \\
&&\times (1-z)^{-1+\eta} U_{G,ucs}(z_{\rm cut}, m, \mu_{ucs}) \tilde{U}_{cus}\bigl[\ln\frac{\mu_{cus}^2}{m^2(1-z)^2}-2\partial_{\eta}\bigr]\frac{e^{-\gamma_E \eta}}{\Gamma(\eta)},\nnb
\eea 
where 
\bea
\mc{M}(\mu_f,\mu_c,\mu_{cs},\mu_{ucs},\mu_{cus}) &=& -2 S_\Gamma(\mu_c,\mu_{cs}) -2 S_\Gamma(\mu_{ucs},\mu_{cus})  -\ln\frac{\mu_c^2}{E_J^2R} a_\Gamma(\mu_c,\mu_{cs}) \nnb\\
&&- \ln\frac{\mu_{ucs}^2}{m^2} a_\Gamma(\mu_{ucs},\mu_{cus}) - \ln z_{\rm cut}^2~a_\Gamma(\mu_{cs},\mu_{cus}) \nnb\\
&& - \frac{C_F}{\beta_0}\left(3\ln\frac{\alpha_s(\mu_f)}{\alpha_s(\mu_c)} +2 \frac{\alpha_s(\mu_{ucs})}{\alpha_s(\mu_{cus})}\right) .
\eea

\begin{figure}[htb] 
	\begin{center}
		\includegraphics[height=7cm]{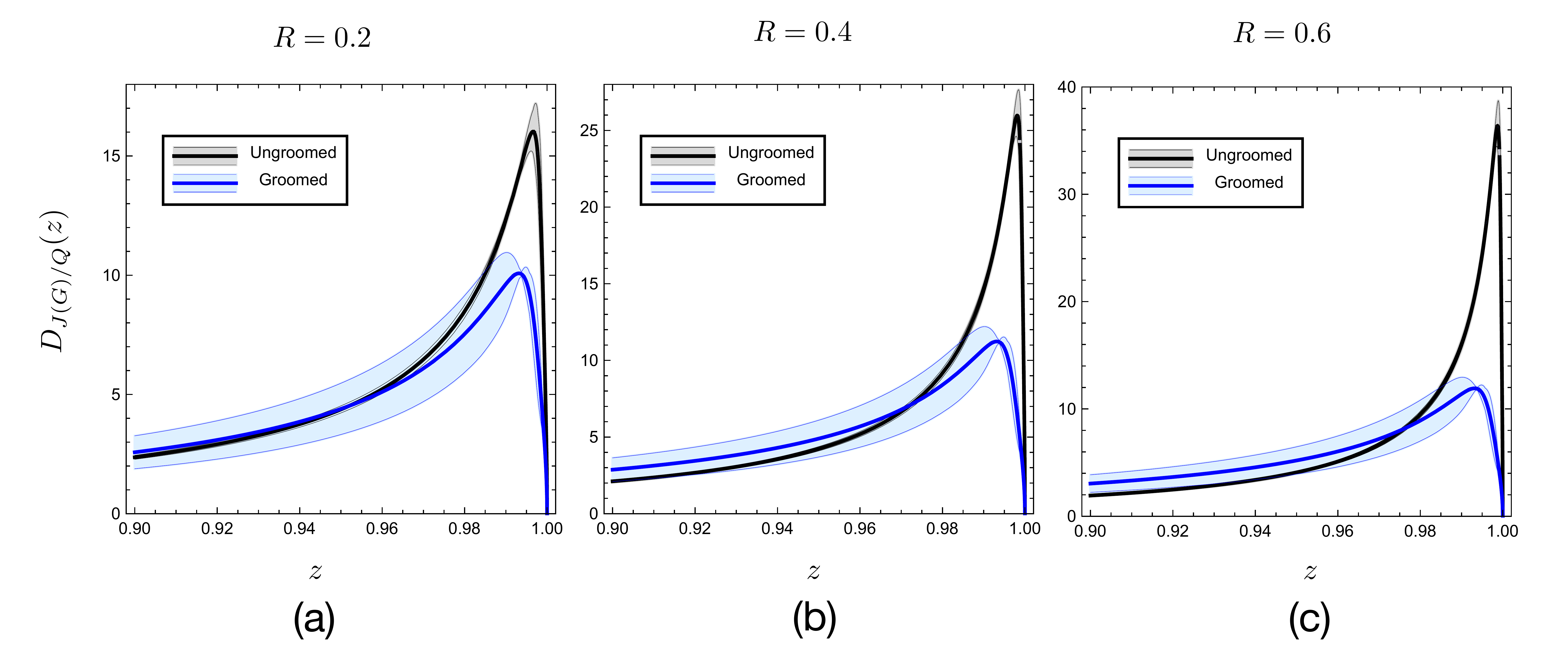}
	\end{center}
	\vspace{-0.8cm}
	\caption{\label{fig:FFJwH}
		FFJ and FFGJ for top quark jets near threshold, with a hierarchy between jet scale (size) and heavy quark mass, i.e., $E_J R \gg m_{t}$.  In this figure, $E_J=2~$TeV and $z_{\rm cut}=0.1$. Panels (a), (b), and (c) show different choices of the jet radius $R=0.2,~0.4,$ and $0.6$, respectively. 
	}
\end{figure}

In Fig.~\ref{fig:FFJwH}, we show the comparison between FFJ and FFGJ for top quark jets when there is a large difference between $E_JR$ and heavy quark mass $m_t$. Here the FFJs and the FFGJs at $\mr{NLL'}$ are plotted with different jet radii; $R=0.2$, $0.4$, and $0.6$, respectively. The error bands are obtained by varying the natural scale $\mu_i$ of each factorized function in Eq.~(\ref{RGFFJG2}) between $\mu_i/2$ and $2\mu_i$ and summing the errors in quadrature.  Since $E_J=2$~TeV, the jet scales $E_JR=0.4$, $0.8$, $1.2$~TeV are well above the top quark mass. As both the FFJ and the FFGJ depend on the factorization scale, we again set it as  $\mu_f=E_J$. In comparison to Fig.~\ref{fig:FFJNoH}(b), due to the large difference between $E_JR$ and $m_t$, grooming effects become more significant. Moreover, for a fixed jet energy, the larger the jet radius is, the greater effect the jet grooming has. The reason is that a larger jet radius means more soft gluons will be enclosed in the jet and consequently more gluons are available for grooming.

\subsection{Cross Sections for Groomed Jets Near Threshold}
Similar to Eq.\,(\ref{eq:xsection}) for ungroomed jet cross sections in  $e^+e^-$ collisions, the factorized form of the groomed jet cross section for $e^+e^-$ collisions can be expressed as 
\begin{equation}
	\label{eq:Gxsection}
	\frac{d \sigma}{d x}=2\sigma_0 \int_{x}^{1} \frac{d z}{z} H_{Q}\left(z ; Q^{2}, \mu\right) D_{J_{G} / Q}\left(\frac{x}{z} ; E_{J} R, m, z_{\rm cut}, \mu\right),
\end{equation}
with the FFJ replaced by the groomed FFJ. In the following, we numerically study the groomed jet cross sections in the limit $(1-z)\ll z_c$ and $m\ll E_JR$, where Eq.\,(\ref{RGFFJ}) for the groomed FFJ is used in Eq.\,(\ref{eq:Gxsection}) for the groomed jet cross section.  The final form of the factorized groomed cross section in the limit $(1-z)\ll z_c$ and $m\ll E_JR$ is
\bea
\frac{1}{\sigma_0}\frac{d\sigma_G}{dx} &=& 2 \exp[\mc{M}(\mu_h,\mu_{hc},\mu_c,\mu_{cs},\mu_{ucs},\mu_{cus})] \nonumber \\
&&\times H(Q,\mu_h) \mc{J}_Q (E_JR,m,\mu_{c}) S_{G,0}\left(z_{\rm cut} E_{J} R, \mu_{cs}\right) U_{G,ucs}\left(z_{\rm cut} m, \mu_{ucs}\right) \nonumber\\
&&\times (1-x)^{-1+\eta} \tilde{J}_{\n} \Bigl[\ln\frac{\mu_{hc}^2}{Q^2(1-x)}-\partial_{\eta}\Bigr] 
\tilde{U}_{cus}\bigl[\ln\frac{\mu_{us}^2}{m^2(1-x)^2}-2\partial_{\eta}\bigr]\frac{e^{-\gamma_E \eta}}{\Gamma(\eta)},
\label{eq:GxsectionwH}
\eea
where $\eta=2a_\Gamma(\mu_{hc},\mu_{cus})$ and the RG evolution factor $\cal M$ is 
\bea
&&\mc{M}(\mu_h,\mu_{hc},\mu_c,\mu_{cs},\mu_{ucs},\mu_{cus})\nonumber\\
&&=4 S_{\Gamma}\left(\mu_{h}, \mu_{hc}\right)-2 S_{\Gamma}\left(\mu_{c}, \mu_{cs}\right)-2 S_{\Gamma}\left(\mu_{ucs}, \mu_{cus}\right)+2 \ln \frac{\mu_{h}^{2}}{Q^{2}} a_{\Gamma}\left(\mu_{h}, \mu_{h c}\right)\nonumber\\
&&\quad -\ln \frac{\mu_{c}^{2}}{E_{J}^{2} R^{2}} a_{\Gamma}\left(\mu_{c}, \mu_{cs}\right)-\ln \frac{\mu_{ucs}^{2}}{m^{2}} a_{\Gamma}\left(\mu_{u cs}, \mu_{cu s}\right)-\ln z_{\rm cut}^{2} a_{\Gamma}\left(\mu_{cs}, \mu_{u cs}\right)\nonumber\\
&&\quad -\frac{C_{F}}{\beta_{0}}\left(3 \ln \frac{\alpha\left(\mu_{h}\right)}{\alpha\left(\mu_{h c}\right)}+3 \ln \frac{\alpha\left(\mu_{h}\right)}{\alpha\left(\mu_{c}\right)}+2 \ln \frac{\alpha\left(\mu_{u cs}\right)}{\alpha\left(\mu_{cu s}\right)}\right).
\eea

\begin{figure}[htb]
	\begin{center}
		\includegraphics[height=7cm]{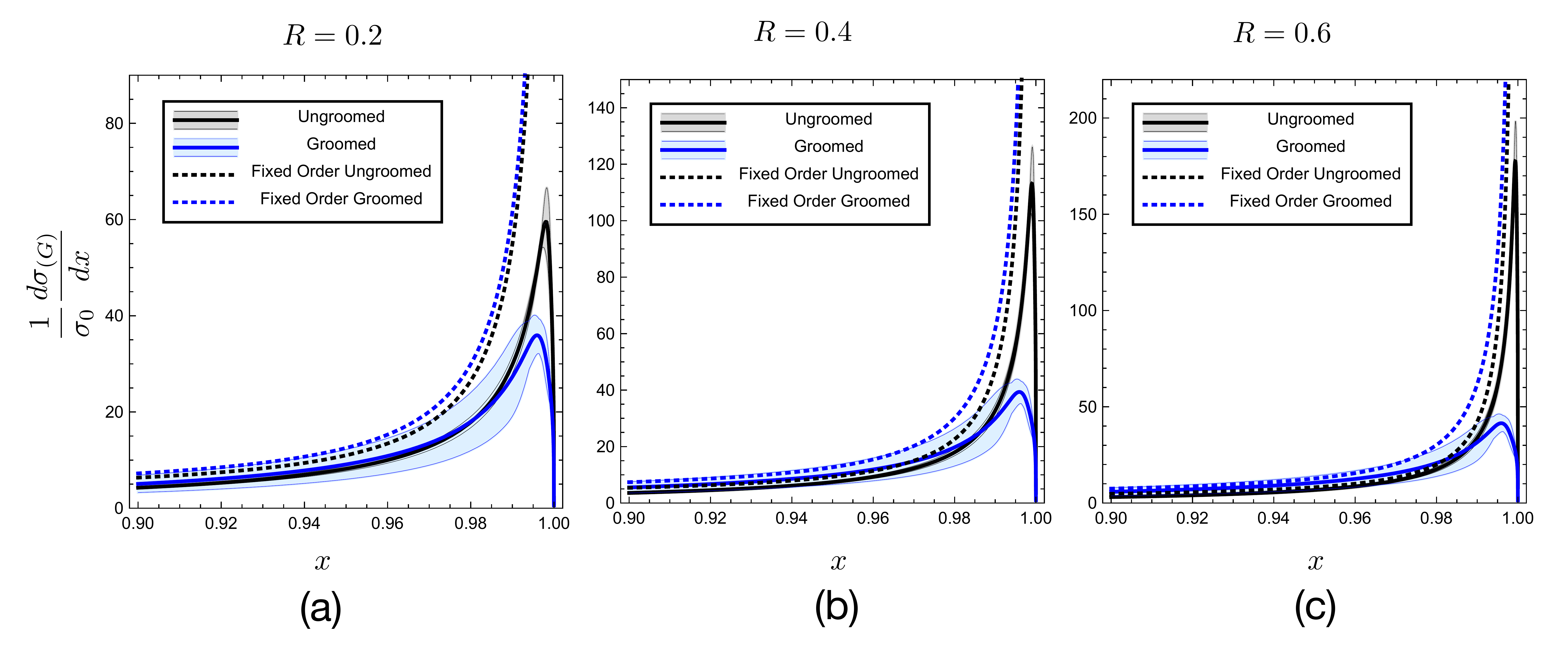}
	\end{center}
	\vspace{-0.8cm}
	\caption{\label{fig:GwH} 
		Comparison between ungroomed and groomed jet cross sections of top quark jets in electron-positron collisions. The black (blue) bands are groomed (ungroomed) jets with resummation with $\mr{NLL'}$ accuracy, and the black (blue) dotted curves are groomed (ungroomed) jets with the fixed order NLO calculation. (a), (b), and (c) correspond to different jet radii, $R = 0.2, 0.4$, and $0.6$, respectively. The center of mass energy is $Q=4~$TeV and the jet energy is $E_J=Q/2 =2~$TeV.
	}
\end{figure}

In Fig.~\ref{fig:GwH}, we show the comparison between ungroomed and groomed jet cross sections for top jet production in electron-positron collisions. The center of mass energy is $Q=4$~TeV, so the top quark jet energy is $2$~TeV, which is the same jet energy that was used of Fig.~\ref{fig:FFJwH}. With this choice, the scale separation $E_JR \gg m_t$ is manifest and the factorized formula Eq.~(\ref{eq:GxsectionwH}) is applicable. The error bands are obtained by varying the natural scale $\mu_i$ of each factorized function in Eq.~(\ref{eq:GxsectionwH}) between $\mu_i/2$ and $2\mu_i$ and summing the errors in quadrature. The groomed jet cross sections share the same features as that of the FFGJs shown in Fig.~\ref{fig:FFJwH}, i.e., the grooming effects are significant near  threshold, and the larger the jet radius, the greater effect of the jet grooming. As we consider jet cross sections for the case $E_JR \sim m_t$, the grooming effects are rather small, similar to Fig.~\ref{fig:FFJNoH}(b).
Also the error bands of the ungroomed and the groomed jet cross sections almost overlap in this case since the errors at the cross section level (i.e., the combination of the errors for many factorized functions) are large. 

The dotted curves in Fig.~\ref{fig:GwH} are the fixed order NLO results. Since the cross section is renormalization scale independent, the dependence on $\mu$ for the fixed order results is only present in $\as(\mu)$, where 
we have chosen $\mu$ to be the hard scale $Q$. In contrast to resummed results (black and blue bands),  the fixed order cross sections become divergent as $x$ approaches $1$, and near the endpoint they are approximated as 
\bea 
\label{fixs} 
\frac{1}{\sigma_0} \Bigl(\frac{d\sigma}{dx}\Bigr)_{\mr{fix}}  &\approx& \frac{\as C_F}{\pi} \frac{1}{1-x} \ln \frac{Q^2}{B^2(1-x)} \sim 
\frac{\as C_F}{\pi} \frac{1}{1-x} \ln \frac{Q^2}{(E_JR)^2(1-x)}, \\
\label{fixsG} 
\frac{1}{\sigma_0} \Bigl(\frac{d\sigma_G}{dx}\Bigr)_{\mr{fix}}  &\approx& \frac{\as C_F}{\pi} \frac{1}{1-x} \ln \frac{Q^2}{m^2(1-x)},
\eea 
where $\sigma$ and $\sigma_G$ are the ungroomed and groomed cross section respectively.  
Fig.~\ref{fig:GwH} also shows that the fixed order results of the groomed cross sections are larger than the ungroomed ones unrealistically, which can be confirmed from Eqs.~\eqref{fixs} and \eqref{fixsG} as well. This also shows that resummation is essential to the formulation of the groomed jet cross section given in this work. 

\section{Conclusion}
\label{conclusion}

In this paper, we studied the process of a heavy quark fragmenting into a jet  in the endpoint region where the jet carries almost all of the energy of the initiating heavy quark, i.e., $z \sim 1$ where $z$ is the jet energy fraction of the fragmenting parton. This analysis employs the heavy quark FFJ, initially introduced in Ref.~\cite{Dai:2018ywt}. (The FFJ was originally introduced for the massless case in Ref.~\cite{Dai:2016hzf} and then studied in the endpoint in Ref.~\cite{Dai:2017dpc}.) We are able to simultaneously resum logarithms of the jet radius $R$ and $1-z$ using Soft Collinear Effective Theory, or more accurately $\mr{SCET_M}$. In the endpoint region, to describe the csoft interactions of the heavy quark, it is useful to match this effective theory onto boosted heavy quark effective theory, which we do in detail. From the resummed result, we show that there are numerically significant corrections to the lowest order result.

One of the advantages of this analysis is that we can use the formalism to investigate heavy quark jet substructure, which we do by calculating the heavy quark fragmentation to a groomed jet, using the soft drop \cite{Larkoski:2014wba} grooming algorithm as an example.  In order to account for the grooming procedure fully, we must introduce a collinear-ultrsoft (cusoft) mode, which is sensitive to the region with $z > 1-\zc$ in the grooming algorithm. 
Then, by considering the heavy quark fragmentation function to a groomed jet in the large $z$ limit, we are able to additionally resum large logarithms of $\zc/(1-z)$. We show that there are again large numerical corrections in the endpoint from this procedure when compared with the analysis on the ungroomed jet. 

A nice feature of this analysis using grooming is the heavy quark mass makes the soft drop algorithm IR finite.
This allows us to perturbatively calculate the corrections and resum the logarithms from grooming except for the extremely close to the endpoint, $1-z \lesssim \Lambda_{\mr{QCD}}/m$. Comparing the sizes of $E_JR'$ and $m$, 
we investigate both the grooming analyses for $E_J R' \sim m$ and for $E_J R'\gg m$. For the $E_J R'\gg m$ kinematics, we refactorize the (original) csoft function introducing the ultracollinear-soft (ucsoft) mode.  With this refactorization we can additionally resum the large logarithms of $E_JR'/m$. 

Finally, as an application, we calculate the rate for $e^+e^-$ collisions to produce a heavy quark jet in the endpoint region, comparing the groomed and the ungroomed results. For specificity, we looked at a hypothetical 4 TeV machine producing top quark jets. We find that the grooming procedure has a sizable effect close to the endpoint. It would certainly be exciting to be able to test this out at a future collider.

\acknowledgments

LD was supported by the Foreign Postdoctoral Fellowship Program of the Israel Academy of Sciences and Humanities, Israeli Science Foundation (ISF) grant \#1635/16, and Binational Science Foundation grant  \#2018722. This work has been performed in the framework of COST Action CA 15213 ``Theory of hot matter and relativistic heavy-ion collisions" (THOR), MSCA RISE 823947 ``Heavy ion collisions: collectivity and precision in saturation physics''  (HI\-EIC) and has received funding from the European Un\-ion's Horizon 2020 research and innovation programm under grant agreement No. 824093. CK is supported by Basic Science Research Program through the 
National Research Foundation of Korea (NRF) funded by the Ministry of Science and ICT (Grants No. NRF-2017R1A2B4010511, No. NRF-2021R1A2C1008906). AL is supported in part by the National Science Foundation under Grant No. PHY-1820760. 

\appendix

\section{Construction of bHQET}
\label{bHQET}

In this appendix we consider a direct construction of the bHQET Lagrangian from $\mr{SCET_M}$. The $\mr{SCET_M}$ Lagrangian is \cite{Leibovich:2003jd}
\bea
\label{SCETM}
\mc{L}_{\mr{SCET_M}}&=&\bar{\xi}_n \Bigl[ n\cdot i\mc{D}+i\fmsl{\mc{D}}_{\perp}\frac{1}{\n\cdot i\mc{D}}i\fmsl{\mc{D}}_{\perp}\Bigr] \nn\xi_n -m^2 \bar{\xi}_n \frac{1}{\n\cdot i\mc{D}} \nn \xi_n \nnb\\
&&+ m\bar{\xi}_n \Bigl[ i\fmsl{\mc{D}}_{\perp},\frac{1}{\n\cdot i\mc{D}}\Bigr]\nn \xi_n.
\eea
We first separate csoft interactions from collinear interactions by writting the covariant derivative as $\mc{D}^{\mu} = D_c^{\mu} + D^{\mu}$, where $D^{\mu}$ is the csoft covariant derivative. 
We then integrate out the collinear modes (gluons). Finally, the SCET massive quark field is matched onto the bHQET field as shown in Eq.~\eqref{Qmatch}.  

The resulting bHQET Lagrangian is 
\bea
\label{bHQET1} 
\mc{L}_{\mr{bHQET}} &=& \frac{v_+}{2} \Bigl\{\bar{h}_n \Bigl(\frac{m}{v_+} + iD_- + i\fmsl{D}_{\perp} \frac{1}{mv_++iD_+} i\fmsl{D}_{\perp} \Bigr) \nn h_n \\
&&~-\bar{h}_n \frac{m^2}{mv_++iD_+}  \nn h_n + \bar{h}_n \Bigl[i\fmsl{D}_{\perp}, \frac{m}{mv_++iD_+}\Bigr]\nn h_n\Bigr\}, \nnb
\eea
where $D_+\equiv \n\cdot D$ and $D_-\equiv n\cdot D$. $v_+=\n\cdot v \sim 2E/m$ is power-counted as $\mc{O}(1/\lambda)$, and the velocity $v^{\mu}$ has been expressed as $(v_+ n^{\mu}+\n^{\mu}/v_+)/2$ with $v_{\perp}$ set to zero. The scaling of the csoft derivative $D^{\mu}$ is $(D_+,D_{\perp},D_-) \sim 2E \eta (1,\lambda,\lambda^2)$. 

Expanding the propagator, 
\be
\frac{1}{mv_++iD_+} = \frac{1}{mv_+} \Bigl[1-\frac{iD_+}{mv_+}+\Bigl(\frac{iD_+}{mv_+}\Bigr)^2 + \mc{O}(\eta^3) \Bigr], 
\ee
we rewrite Eq.~\eqref{bHQET1} to $\mc{O}(\eta)$ as 
\bea
\label{bHQET2}
\mc{L}_{\mr{bHQET}} &=& \frac{v_+}{2} \bar{h}_n \Bigl( iD_- + \frac{iD_+}{v_+^2}\Bigr)\nn h_n \\
&&+\frac{1}{2m}\bar{h}_n \Bigl( i\fmsl{D}_{\perp} i\fmsl{D}_{\perp}- \frac{(iD_+)^2}{v_+^2} - \Bigl[i\fmsl{D}_{\perp},\frac{iD_+}{v_+}\Bigr]\Bigr)\nn h_n. \nnb
\eea 
The term in the first line is the leading Lagrangian, which can be re-expressed as 
\be
\label{leadingL}
\mc{L}_{\mr{bHQET}}^{(0)} = \bar{h}_n v\cdot iD \nn h_n. 
\ee

Using leading equation of motion from \eqref{leadingL}, we have the relations
\be 
\label{rel}
\frac{iD_+}{v_+} \nn h_n = -v_+ iD_- \nn h_n,~~~
\bar{h}_n \nn \frac{i\overleftarrow{D}_+}{v_+}  = -v_+ \bar{h}_n \nn i\overleftarrow{D}_-,
\ee
where $i\overleftarrow{D}^{\mu} = -i\overleftarrow{\partial}^{\mu} +gA_{cs}^{\mu}$.
Applying Eq.~\eqref{rel}, we can further simplify the subleading terms in Eq.~\eqref{bHQET2}, 
\bea
&&\bar{h}_n \Bigl( i\fmsl{D}_{\perp} i\fmsl{D}_{\perp}- \frac{(iD_+)^2}{v_+^2} - \Bigl[i\fmsl{D}_{\perp},\frac{iD_+}{v_+}\Bigr]\Bigr)\nn h_n \nnb \\
&&~~~= \bar{h}_n \Bigl(iD_+ iD_- +i\fmsl{D}_{\perp} i\fmsl{D}_{\perp} - \Bigl[i\fmsl{D}_{\perp},\frac{iD_+}{v_+}\Bigr]\Bigr)\nn h_n \nnb \\
\label{subl}
&&~~~=\bar{h}_n \left[(iD)^2 +g\left(\frac{\sigma^{\mu\nu}_{\perp}}{2}+\frac{i\n^{\mu}\gamma_{\perp}^{\nu}}{v_+}\right)G_{\mu\nu} \right]\nn h_n,  
\eea
where $\sigma^{\mu\nu}_{\perp} = i[\gamma_{\perp}^{\mu},\gamma_{\perp}^{\nu}]/2$, and we used the relation $[iD^{\mu},iD^{\nu}] = ig G^{\mu\nu}$. 

Finally, the bHQET Lagrangian to $\mc{O}(\eta)$ is given by  
\bea
\label{bHQET3}
\mc{L}_{\mr{bHQET}} = \bar{h}_n v\cdot iD \nn h_n
+\frac{1}{2m} \bar{h}_n \left[(iD)^2 +g\left(\frac{\sigma^{\mu\nu}_{\perp}}{2}+\frac{i\n^{\mu}\gamma_{\perp}^{\nu}}{v_+}\right)G_{\mu\nu} \right]\nn h_n. 
\eea
The subleading kinematic term $(iD)^2$ is connected to the leading Lagrangian by reparameterization invariance, since
$v^{\mu} \to v^{\mu} + iD^{\mu}/m$~\cite{Luke:1992cs}. So this term has no higher order corrections in $\as$ nor any nontrivial renormalization. On the other hand, we expect that the chromomagnetic term with $G_{\mu\nu}$ to have  higher order corrections and  will need renormalization similarly to standard HQET.  

Since the bHQET field $h_n$ has a different spin property from the standard HQET field $h_v$, the relation between the two fields needs to be investigated. In order to do so, we first consider the reparameterization invariant (RPI) combination for both  fields. Note that in order for the combination to fully preserve the reparameterization symmetries, it must reproduce the form of the full theory quark field (with a large phase removed). Therefore the combinations for both  fields should be equal to each other. 

Since $h_n$ has been matched from $\xi_n$ in $\mr{SCET_M}$, we can find the form in bHQET from $\mr{SCET_M}$. In $\mr{SCET_M}$ the RPI form is given by~\cite{Chay:2005ck,Chay:2002vy,Manohar:2002fd}
\be 
\label{SCETRPI}
\xi_n+ \frac{1}{\n\cdot i\mc{D}}(i\fmsl{\mc{D}}+m)\nn \xi_n.
\ee
To this, if we apply the matching relation, Eq.~\eqref{Qmatch}, the RPI combination of the bHQET field is
\be
\label{bHQETRPI}
\sqrt{\frac{v_+}{2}} \left[h_n + \frac{1}{mv_++iD_+}\left(i\fmsl{D}_{\perp}+m\right)\nn h_n\right].
\ee 
This combination should be equal to the standard HQET combination, given by 
\be
\frac{1+\fms{v}}{2} h_v + \frac{1-\fms{v}}{2}H_v, 
\ee
where $H_v$ satisfies $\fms{v}H_v = -H_v$ and is power-suppressed by $1/m$ when compared with $h_v$. 

Therefore, applying the projection $(1+\fms{v})/2$ to Eq.~\eqref{bHQETRPI}, we have the following relation 
\be
\label{rel0}
h_v = \frac{1+\fms{v}}{2} \sqrt{\frac{v_+}{2}} \left(1 + \frac{1}{mv_++iD_+}(i\fmsl{D}_{\perp}+m)\nn \right) h_n,
\ee
where $v$ has been given in Eq.~\eqref{velo}. Expanding to $\mc{O}(\eta^2)$, we obtain
\be
\label{rel1} 
h_v = \sqrt{\frac{v_+}{2}}\left(1+\frac{1}{v_+} \nn\right) \left[ h_n 
-\frac{1}{2m} \left(i\fmsl{D}_{\perp}+\frac{iD_+}{v_+} \right) h_n\right], 
\ee
where the second term in the square bracket is suppressed by $\eta$. Hence, to leading order $\eta$, $h_n$ is given by  
\be
\label{rel2} 
h_n = \sqrt{\frac{2}{v_+}}\Bigl(1-\frac{1}{v_+} \nn\Bigr) h_v. 
\ee

Substituting Eq.~\eqref{rel2} into Eq.~\eqref{leadingL}, we reproduce the leading HQET Lagrangian as 
\be 
\bar{h}_n v\cdot iD \nn h_n = \frac{2}{v_+} \bar{h}_v v\cdot iD \nn h_v = \bar{h}_v v\cdot iD  h_v. 
\ee
where the second equality is obtained from the relation
\be 
\frac{2}{v_+} \bar{h}_v v\cdot iD \nn \fms{v} h_v = 2 \bar{h} _v v\cdot iD  h_v - \frac{2}{v_+} \bar{h}_v \fms{v} v\cdot iD \nn  h_v.
\ee
The subleading Lagrangian at LO in $\as$ for the standard HQET is given by~\cite{Falk:1990pz}
\be
\label{subHQET}
\mc{L}_{\mr{HQET}}^{(1)} = -\frac{1}{2m} \bar{h}_v \fmsl{D}\fmsl{D}~h_v
= \frac{1}{2m} \Bigl[\bar{h}_v (iD)^2 h_v + \frac{g}{2} \bar{h}_v \sigma^{\mu\nu}G_{\mu\nu} h_v \Bigr], 
\ee
where the chromomagnetic operator containing $G_{\mu\nu}$ has nonzero corrections at higher order in $\as$. 
From Eq.~\eqref{subHQET}, it is straightforward to obtain Eq.~\eqref{bHQET3} using the leading relation in Eq.~\eqref{rel1}.

\section{Profile Functions}
\label{sec:profile}
\begin{figure}[htb]
	\begin{center}
		\includegraphics[height=7cm]{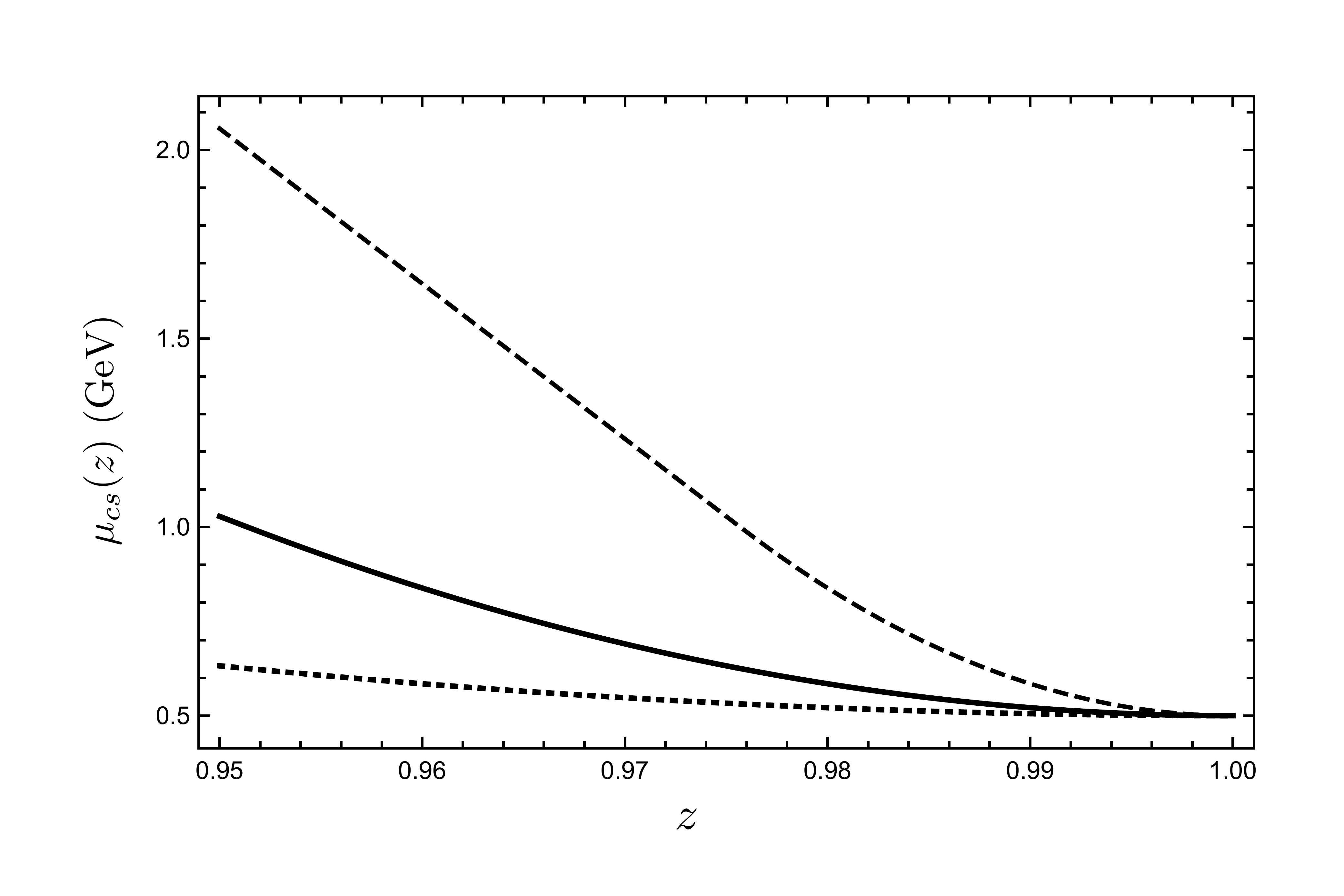}
	\end{center}
	\vspace{-0.8cm}
	\caption{\label{fig:Profile}\,The profile function for the $z$-dependent csoft scale as appears in Eq.~(\ref{fact2}) where $B=\sqrt{E_J^2R^2+m^2}$ with $E_J=50$\,GeV, $m=4.8$\,GeV, and $R=0.4$. Here $\mu_{min}$ is set to be $0.5$\,GeV. The solid, dashed, and dotted curves correspond to $\mu_0=B$, $2B$, and $B/2$, respectively.
	}
\end{figure}

For a $z$-dependent scale like the csoft scale $\mu_{cs}=B(1-z)$ that appears in Eq.~(\ref{fact2}), the scale becomes non-perturbative as $z$ approaches $1$. To enable numerical evaluations, we freeze such $z$-dependent scales to a certain value $\mu_{min}$ using the following equation,
\begin{equation}
	\mu_{\rm profile}(z)=\left\{
	\begin{array}{l}
		\mu_0(1-z),~ \text{for~} z<x; \\
		\mu_{min}+a \mu_0(1-z)^2,~ \text{for } z\ge x.
	\end{array}
\right.
\end{equation}
where $\mu_0$ is $\mu_0=B=\sqrt{E_J^2R^2+m^2}$ for the csoft scale, or $\mu_0=m$ for the usoft scale, and $x$ and $a$ are fixed by requiring $\mu_{\rm profile}(z)$ to be continuous and smooth (differentiable) at $x$. When the scale variation for a relevant function  is evaluated, $\mu_0$ varies between $\mu_0/2$ and $2\mu_0$. As an example, the profile functions for the csoft scale is shown in Fig.\,\ref{fig:Profile}.




\bibliographystyle{JHEP1}
\bibliography{Jet}



\end{document}